\documentclass[prd,nofootinbib,showpacs,floatfix,preprintnumbers]{revtex4}
\usepackage{epsfig}
\usepackage[dvips]{color}
\usepackage{graphicx}
\input epsf

\def\be{\begin{equation}}
\def\ee{\end{equation}}
\def\bea{\begin{eqnarray}}
\def\eea{\end{eqnarray}}
\def\d{{\rm d}}
\def\lsim{\raise0.3ex\hbox{$\;<$\kern-0.75em\raise-1.1ex\hbox{$\sim\;$}}}
\def\gsim{\raise0.3ex\hbox{$\;>$\kern-0.75em\raise-1.1ex\hbox{$\sim\;$}}}

\def\theta{\vartheta}

\definecolor{Black}{named}{Black}
\definecolor{Red}{named}{Red}

\newcommand{\bw}{\begin{widetext}}
\newcommand{\ew}{\end{widetext}}
\begin{document}
\def\d{{\rm d}}
\title{Disentangling neutrino-nucleon cross section and high energy neutrino flux with a km$^3$ neutrino telescope}
\author{E.~Borriello$^{1,2}$, A.~Cuoco$^{3}$, G.~Mangano$^{1}$, G.~Miele$^{1,2}$, S.~Pastor$^{2}$,
O.~Pisanti$^{1}$, and P.~D.~Serpico$^{4}$}
\affiliation{ $^1$Universit\`{a} ``Federico II" Dipartimento di
Scienze Fisiche, \& INFN Sezione di Napoli,  Napoli, Italy}
\affiliation{$^2$ AHEP Group, Institut de F\'{\i}sica Corpuscular, CSIC/Universitat de Val\`{e}ncia,
Apt. 22085, 46071 Val\`{e}ncia, Spain}
\affiliation{$^3$ Department of Physics and Astronomy, University of
Aarhus, Ny Munkegade, 8000 Aarhus, Denmark,}
\affiliation{$^4$ Center for Particle Astrophysics, Fermi National Accelerator Laboratory, Batavia, IL 60510-0500  USA}
\date{\today}
\begin{abstract}
The energy--zenith angular event distribution in a neutrino
telescope provides a unique tool to determine at the same time the
neutrino-nucleon cross section at extreme kinematical regions, and
the high energy neutrino flux. By using a simple parametrization
for fluxes and cross sections, we present a sensitivity analysis
for the case of a km$^3$ neutrino telescope. In particular, we
consider the specific case of an under-water Mediterranean
telescope placed at the NEMO site, although most of our results
also apply to an under-ice detector such as IceCube. We determine
the sensitivity to departures from standard values of the cross
sections above 1 PeV which can be probed independently from an
a-priori knowledge of the normalization and energy dependence of
the flux. We also stress that the capability to tag downgoing
neutrino showers in the PeV range against the cosmic ray induced
background of penetrating muons appears to be a crucial
requirement to derive meaningful constraints on the cross section.
\end{abstract}
\pacs{13.15.+g, %Neutrino interactions
95.55.Vj ,      %Neutrino, muon, pion, and other elementary particle detectors; cosmic ray detectors
95.85.Ry        %Neutrino, muon, pion, and other elementary particles; cosmic rays
}
\maketitle

\preprint{DSF-36-2007, FERMILAB-PUB-07-582-A, IFIC/07-60}

%%%%%%%%%%%%%%%%%%%%%%%%%%%%%%%%%%%%%%%%%%
\section{Introduction}\label{introduction}
%%%%%%%%%%%%%%%%%%%%%%%%%%%%%%%%%%%%%%%%%%
%
High energy neutrino astronomy is one of the most promising
research lines in astroparticle physics. Neutrinos are in fact one
of the main components of the cosmic radiation in the high energy
regime, and although their fluxes are uncertain and depend on the
specific production process, their detection would provide
valuable information concerning the sources and the acceleration
mechanism in extreme astrophysical environments. Similarly to
photons and unlike cosmic rays, they keep directional information
which can be used to perform astronomy. Differently from gamma
rays, they are emitted only in hadronic processes and travel
unimpeded through cosmological distances well above the TeV
energy. Moreover, due to the peculiar charge assignment of
neutrinos in the electroweak standard model, they may be the
particles most sensitive to exotic physics. An intriguing
possibility is that a measurement of their cross section on
ordinary matter in extreme and still unexplored kinematical
regions might unveil new physics (see e.g.
\cite{Anchordoqui:2005is}).

From the experimental point of view, after the first generation of
telescopes has proved the viability of the Cerenkov detection
technique under deep water (\verb"Baikal" \cite{Balkanov:1999up})
and ice (\verb"AMANDA" \cite{Ahrens:2002gq}) by detecting
atmospheric neutrinos, one is probably on the verge of the first
detections at the \verb"IceCube" \cite{Ahrens:2003ix} telescope,
being completed at the South Pole, and possibly at the smaller
\verb"ANTARES" \cite{antares} telescope under construction in the
Mediterranean. Additionally, \verb"ANTARES" as well as
\verb"NESTOR" \cite{nestor} and \verb"NEMO" \cite{nemo} are
involved in R\&D projects aimed at the construction of a km$^3$
Neutrino Telescope (NT) in the deep water of the Mediterranean
sea, coordinated in the European network \verb"KM3NeT"
\cite{km3net}.

In the present paper we address in a more quantitative way the
question of how well a km$^3$ NT performs in the simultaneous
determination of the neutrino flux and the neutrino-nucleon cross
section by using as observable the energy--zenith angular event
distribution.  Note that, unless the spectrum is known a priori,
for a NT surrounded by an isotropic medium the cross section and
the flux would be completely degenerate. However, under the sole
assumption that the incoming spectrum is isotropic, the different
opacity of the Earth in different directions --- which becomes
relevant above $\sim$100 TeV --- allows one to disentangle the two
observables, as it has been discussed before e.g. in
\cite{Hooper:2002yq,Hussain:2006wg,Anchordoqui:2005pn}. In other
words, a NT is only sensitive to the product of cross section and
flux, but the incoming isotropic flux is shielded (via a
non-trivial function of the cross section) in a
direction-dependent way, which makes possible to break the
degeneracy by means of directional information. Here we study this
property using a parametric and phenomenological approach, rather
than illustrating it with a few discrete cases from models
proposed in the literature.

In particular, we consider the specific case of an under-water
Mediterranean telescope, although most of our results also apply
to an under-ice detector such as IceCube. Since the three
different proposed sites for the under-water km$^3$ telescopes
have shown event rate differences of the order of 20\%
\cite{Cuoco}, for the sake of brevity we report the results of our
analysis for the \verb"NEMO" site only, which shows intermediate
performances.

This paper is structured as follows. In Sec. \ref{parameteriz} we
introduce the parameterization used for the flux and cross
section. In Sec.~\ref{formalism} we describe the formalism used,
while our results are reported in Sec. \ref{results}. Finally, we
conclude in Sec. \ref{conclusions}.

%%%%%%%%%%%%%%%%%%%%%%%%%%%%%%%%%%%%%%%%%%%%%%%%%%%%%%%%%%%%%%%%%%%%%%%%
\section{Parameterization of Flux and cross sections}\label{parameteriz}
%%%%%%%%%%%%%%%%%%%%%%%%%%%%%%%%%%%%%%%%%%%%%%%%%%%%%%%%%%%%%%%%%%%%%%%%
%
Many astrophysical objects are expected to produce a flux of high
energy neutrinos, possibly with energies higher than the highest
energy cosmic rays ever detected, of the order of 10$^{11}\,$GeV.
The center of mass energy available in a neutrino-nucleon
collision for a neutrino with laboratory energy $E_\nu$ is
$\sqrt{s} \simeq \sqrt{2E_\nu\,m_N}\simeq
1.4\,$TeV$\sqrt{E_\nu/{\rm PeV}}$, $m_N$ being the nucleon mass.
Already at PeV incident energies, this is well above the
electro-weak symmetry scale of $\cal{O}$(100) GeV and may thus
reveal the onset of physics beyond the standard model\footnote{Of
course, instruments looking at the highest energy events, say
above 10$^{10}\,$GeV, have greater chances to detect significant
departures from the standard model behavior, although the limited
statistics likely requires future space based fluorescence
detector (for an analysis of this case, see for example
\cite{PalomaresRuiz:2005xw}).}.  Since neutrinos only respond to
weak interactions, even weak effects may be revealed by a study of
the neutrino-nucleon cross sections at NTs. One example is
provided by low-scale quantum gravity models, see e.g.
\cite{AlvarezMuniz:2001mk}.

Unfortunately, there are two major problems to be addressed in
such a program: the low statistics expected at NTs, which requires
huge volumes and long operation times, and the ignorance of the
diffuse astrophysical spectrum. At the range of energy
$E_\nu\sim$PeV which starts to be interesting for particle physics
constraints, the natural target of NTs are diffuse fluxes rather
than point-like sources. On the basis of existing predictions for
extragalactic fluxes,  we limit our analysis  to energies above
$10^{-0.5}\,$PeV, where  isotropic diffuse extragalactic fluxes
are expected to dominate over the steeper atmospheric neutrino
flux. Around PeV energies, the atmospheric neutrino event rates
are very small even in a km$^3$ telescope and the only remaining
background is given by penetrating muons from extensive showers
due to cosmic ray events, which may require an active veto to be
disentangled from the signal. Given the observed behavior of
cosmic ray and gamma ray spectra, it is reasonable to parameterize
the neutrino spectra as a power law or, more generally, as a
broken power-law. The latter parameterization would be probably
needed when considering the flux over several decades of energy.
However, for the following discussion most of the useful events
will have primary energy $E_\nu< 10\,$PeV, and the single
power-law approximation seems reasonable for a single decade in
energy. The spectral index in Fermi-like acceleration mechanisms
is of the order of $-2$ or slightly steeper, but this observable
is expected to be determined by the data as well. In the
following, we parameterize the neutrino flux (per flavor, summing
neutrinos and anti-neutrinos) as
\begin{equation}
\phi_\nu(E_\nu) \equiv\frac{\d^2 \Phi_\nu}{\d E_\nu\d \Omega}(E_\nu) = 1.3 \cdot
10^{-20} \, C~ \left(\frac{E_\nu}{1\,
\textrm{PeV}}\right)^{-2\, D} \textrm{GeV}^{-1} \textrm{cm}^{-2}
\textrm{s}^{-1} \textrm{sr}^{-1}, \label{flux}
\end{equation}
where $C$ and $D$ are normalization and flux steepness free
parameters, and for $C = D = 1$   one recovers the benchmark case
of a Waxman-Bahcall flux \cite{Waxman:1998yy,Anchordoqui:2005is}.

Concerning the cross sections, we adopt a simple parameterization
as a broken power-law where each energy interval (fixed a priori)
has its own slope, and continuity is imposed at the boundaries.
For example, the charged current (CC) cross section is given by
\begin{eqnarray} \frac{\sigma_{CC}^{\nu N}}{10^{-33} \,
\textrm{cm}^{2}} = \left\{
\begin{array}{lcl}
0.344 \left(\frac{E_\nu}{E_1}\right)^{0.492\,A} &
&E_1 \leq E_\nu \leq E_2\\
0.344
\left(\frac{E_2}{E_1}\right)^{0.492\,A}\left(\frac{E_\nu}{E_2}\right)^{0.492\,B}
& &E_{\nu} > E_2
\end{array}
\right. ,\label{CC}
\end{eqnarray}
where we assume $E_1=10^{-0.5}\,$PeV , $E_2=1\,$PeV and, for
$A=1$, the expression for the first energy bin reduces to the
standard value reported in \cite{Gandhi:1998ri}. An analogous
parameterization is used for the neutral current (NC) cross
section,
\begin{eqnarray} \frac{\sigma_{NC}^{\nu N}}{0.418\cdot 10^{-33} \,
\textrm{cm}^{2}} = \left\{
\begin{array}{lcl}
0.344 \left(\frac{E_\nu}{E_1}\right)^{0.492\,A'} &
&E_1 \leq E_\nu \leq E_2\\
0.344
\left(\frac{E_2}{E_1}\right)^{0.492\,A'}\left(\frac{E_\nu}{E_2}\right)^{0.492\,B'}
& &E_{\nu} > E_2
\end{array}
\right. .\label{NC}
\end{eqnarray}
In the following, we shall consider both cases where CC and NC
cross sections change proportionally to each other (i.e., $A=A'$
and $B=B'$) and the case where only $\sigma_{\rm NC}$ is affected
($A=B=1$ while $A'$ and $B'$ are set free). The former possibility
mimics the scenario, for example, where the current extrapolation
of parton distribution functions is wrong. It is more likely,
though, that ``truly exotic'' new physics would manifest itself
only in NC events; the latter case is a toy model for this class
of situations. There are also cases where new physics manifests
additionally in changes of the inelasticity of the cross sections,
but we leave them outside the range of possibilities explored in
this analysis: so, we shall make standard assumptions on the
inelasticity of the collisions, as reported in more detail in the
next Section.

%%%%%%%%%%%%%%%%%%%%%%%%%%%%%%%%%%%%%%%%%
\section{The formalism} \label{formalism}
\begin{figure}
\begin{tabular}{cc}
\epsfig{file=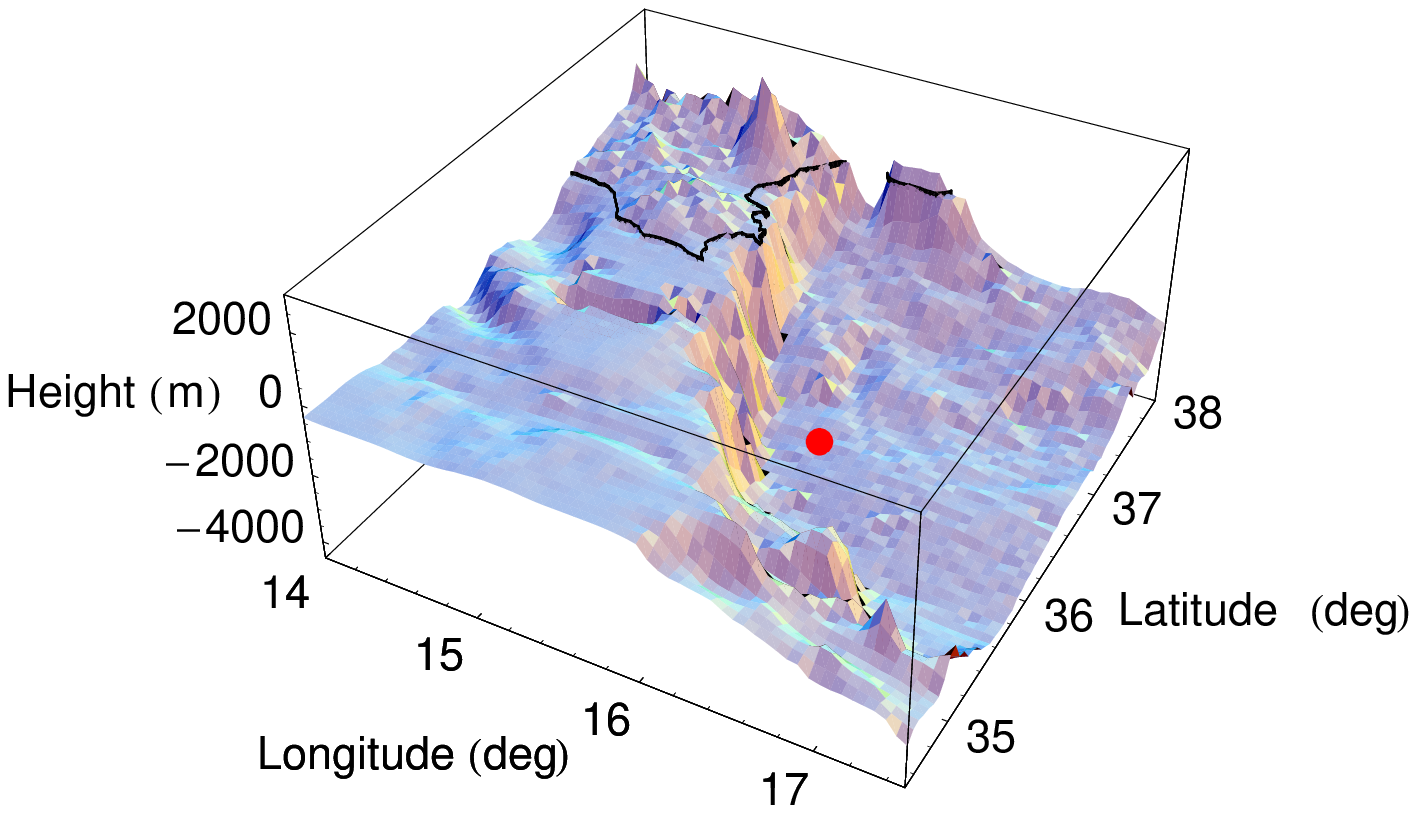,width=8cm} &
\epsfig{file=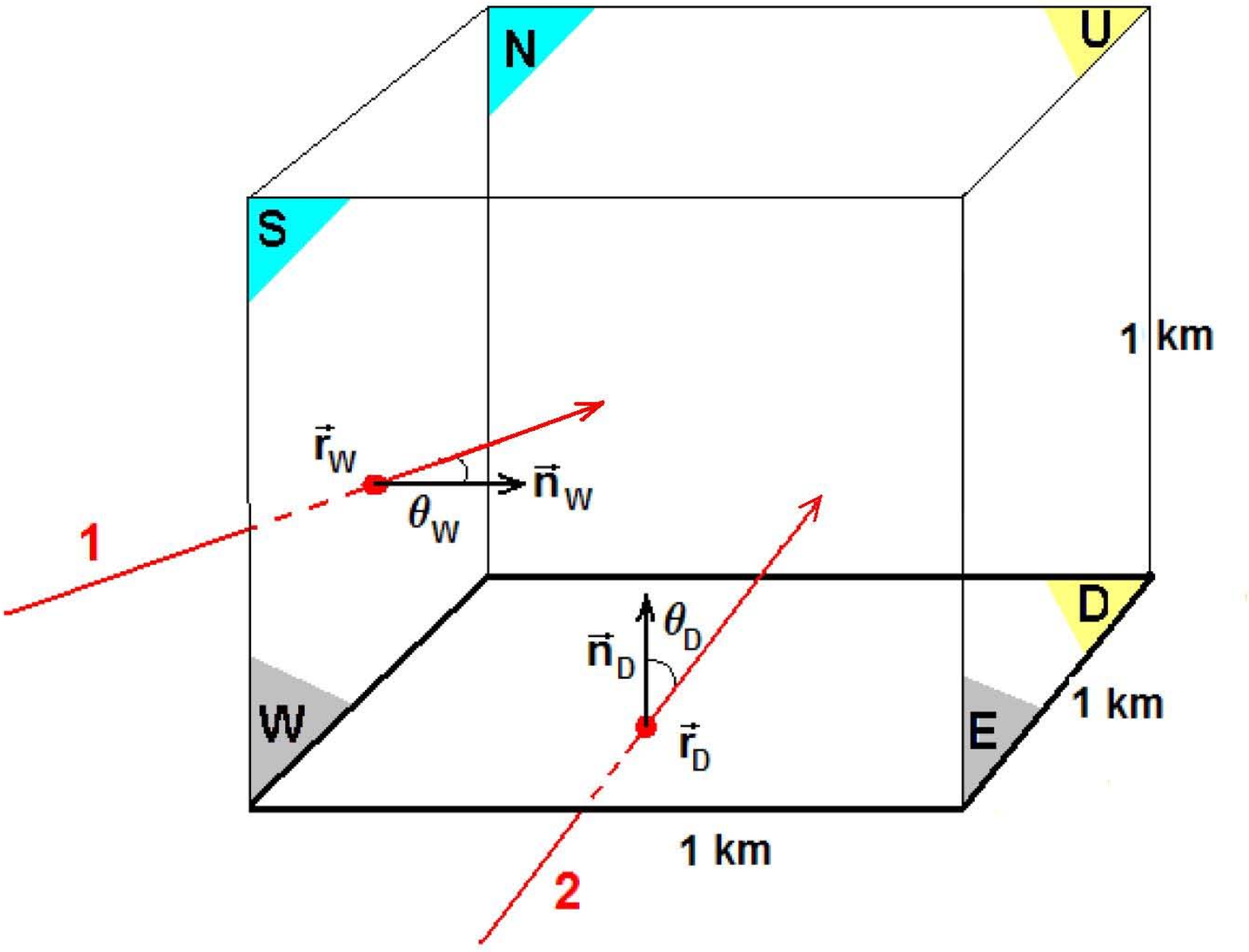,width=6cm}
\end{tabular}
\caption{On the left side is reported the surface profile of the
area nearby the NEMO site, with the red spot giving the NT
location. On the right, we show the assumed NT fiducial volume and
the related notations (see also Ref. \cite{Cuoco}).} \label{NT}
\end{figure}

%%%%%%%%%%%%%%%%%%%%%%%%%%%%%%%%%%%%%%%%%
%
To perform our forecast, we follow the formalism outlined in
\cite{Cuoco}. We generate neutrino tracks isotropically outside
the Earth and assume flavor equipartition due to neutrino
oscillations and pion production dominance \cite{Learned:1994wg}.
We place then a km$^3$ NT at the \verb"NEMO" site, see left panel
of Fig. \ref{NT}. For simplicity we assume the telescope {\it
fiducial} volume as a cube bounded by six lateral surfaces,
labeled D, U, S, N, W, and E for the Down, Up, South, North, West,
and East side, respectively (see right panel of Fig. \ref{NT}).

Despite the different behavior of the produced tau leptons with
respect to muons in terms of energy loss and decay length, both
$\nu_\mu$ and $\nu_{\tau}$ event detection rates are sensitive to
the matter distribution in the neighborhood of the NT site. In
principle, the elevation profile of the Earth surface around the
telescope may be relevant, similarly to what is known for
Earth-skimming ultra-high energy $\nu_{\tau}$'s at extensive air
shower detectors, also investigated by some of the present authors
\cite{Miele} for the Pierre Auger Observatory \cite{Abraham} by
using a Digital Elevation Map (DEM) of the site. In Ref.
\cite{Cuoco} the DEM's of the under-water Earth surface provided
by the Global Relief Data survey (ETOPO2) \cite{ETOPO2} were used
to estimate the effective aperture for $\nu_\tau$ and $\nu_\mu$
detection of a km$^3$ NT in the Mediterranean sea placed at any of
the three locations proposed by the \verb"ANTARES", \verb"NEMO"
and \verb"NESTOR" collaborations. It was found that the effect of
the profile on the total number of events is negligible, although
it may be important in the differential number of events from
different directions, provided one has an excellent angular resolution.
In the present analysis we take into account
the underwater surface profile as done in Ref. \cite{Cuoco},
although for the present results it plays a minor role. Moreover
we also take into account  the radial density profile of the Earth
(reported, for example, in Ref. \cite{Gandhi96}).

Neutrino telescopes were originally thought mostly as $\nu_{\mu}$
detectors, but they have some sensitivity to all neutrino flavors
\cite{Beacom:2003nh}, although this does not trivially translate
into a flavor tagging sensitivity. In particular, their
capability as $\nu_{\tau}$ detectors has become a hot topic in
view of the fact that flavor neutrino oscillations lead to nearly
equal astrophysical fluxes for the three neutrino flavors. While
electron neutrinos only induce shower events inside the fiducial
volume, muon and very high energy tau leptons produced in charged
current neutrino interactions create tracks which may be detected
even when originating far from the instrumented volume. In the
following, we consider as experimental observables:
\begin{itemize}
\item[(i)]  The energy deposited in the detector, $\Delta E$.
\item[(ii)] The topology of the event, namely if it is a shower
or a track event.
\item[(iii)] Some (at least loose) information on the incoming
direction of the event.
\end{itemize}
In turn, these observables depend on some detector-dependent
parameters like the optical properties of the medium, the geometry
of the strings and the position of the optical modules on it, the
efficiencies, etc. Here we take the simplified approach to
consider a (muon or tau) track detected whenever the charged
lepton decay length is longer than the length of the intersection
of the trajectory with  the  instrumented cubic volume. Any other
neutral current event which forms inside the instrumented volume,
or charged current event which does not fulfill the previous
condition is classified as a shower. Although the actual criteria
in a realistic experiment will differ, we expect this
simplification to catch the physics of the problem. We shall
comment on the impact of dropping the assumption of topological
discrimination in Sec. \ref{results}.

Concerning the energy deposited in the detector, for our purposes
it is sufficient to consider two bins only: a low-energy one
(denoted LE) corresponding to $10^{-0.5} \leq \Delta E/{\rm PeV}
\leq 1$ and a high-energy one (denoted HE) for $\Delta E/{\rm PeV}
>1$. Note that we choose the lower bounds of the two bins equal to
the quantities $E_1$ and $E_2$ in Eq.s (\ref{CC}) and (\ref{NC}).
Of course, this is a conventional choice dictated by simplicity
and others are possible,  although a posteriori it reveals to be a
quite reasonable one. In order to calculate the energy released in
the detector, one must also know the fraction of invisible energy
and the energy loss mechanism of charged leptons. The fraction of
energy {\it not} carried by the outgoing lepton and thus released
to the struck nucleon is the inelasticity, $y$. At these energies
$y$ is basically independent of the flavor and is the same for
$\nu$ and $\bar\nu$, decreasing from $\sim 0.3$ at $E_\nu\simeq
E_1$ to $\sim0.2$ at the highest energies of interest
\cite{Gandhi96}.  If the interaction happens inside the detector
(contained event) the energy $y E_\nu$ is considered observed,
since most of the energy in the hadronic shower generated by the
struck hadron is released locally, virtually appearing as a
point-like source\footnote{Even the highest energy showers
penetrate water or ice less than $\sim10$ m, a distance short
compared to any reasonable  spacing of the photomultiplier tubes
(PMTs). The radius over which PMT signals are produced is
$\sim$250 m for a 1 PeV shower; this radius grows or decreases by
approximately 50 m with every decade of shower energy
\cite{Halzen:2002pg}.}. For NC events, this is the only energy
release, since the outgoing neutrino is invisible. For CC events,
all of the remaining energy $(1-y)\,E_\nu$  is released locally
for electron contained or other shower events, while for muon or
tau tracks the energy losses are calculated as in
\cite{Aramo,Dutta2}. Namely, the differential energy loss of the
$\tau$ leptons per unit of length in an underwater NT can be
simply taken as $\d E_\tau/\d \lambda=- \beta_\tau \, E_\tau
\varrho_w$, with $\beta_\tau = 0.71 \times 10^{-6}$ cm$^2$
g$^{-1}$ and $\varrho_w$ denoting the water density. Analogously,
for muons one just needs to replace $\beta_\tau$ with the
corresponding value, $\beta_\mu=0.58 \times 10^{-5}$ cm$^2$
g$^{-1}$.

\begin{figure}[t]
\begin{tabular}{cc}
\epsfig{file=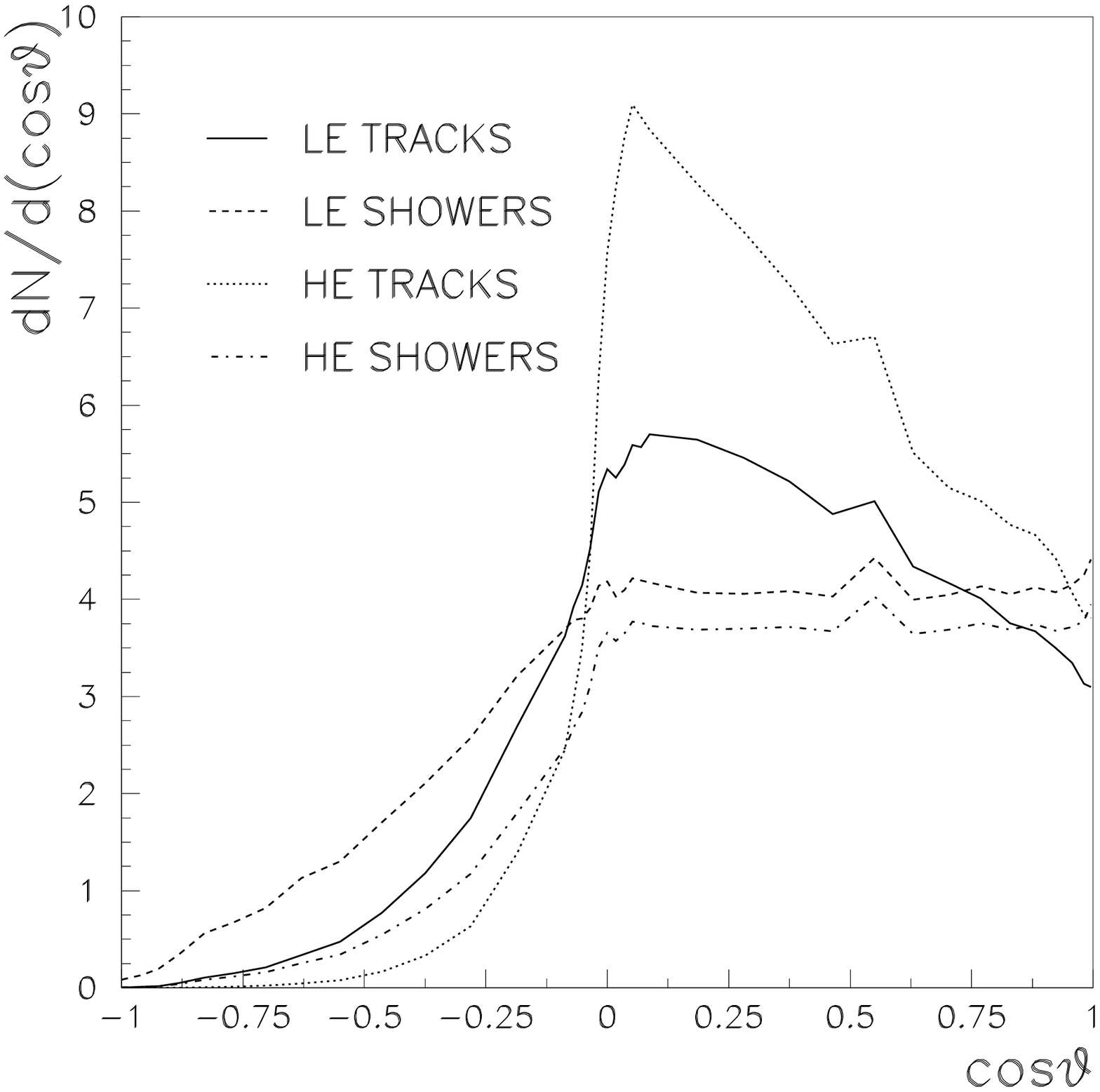,width=6cm} &
\epsfig{file=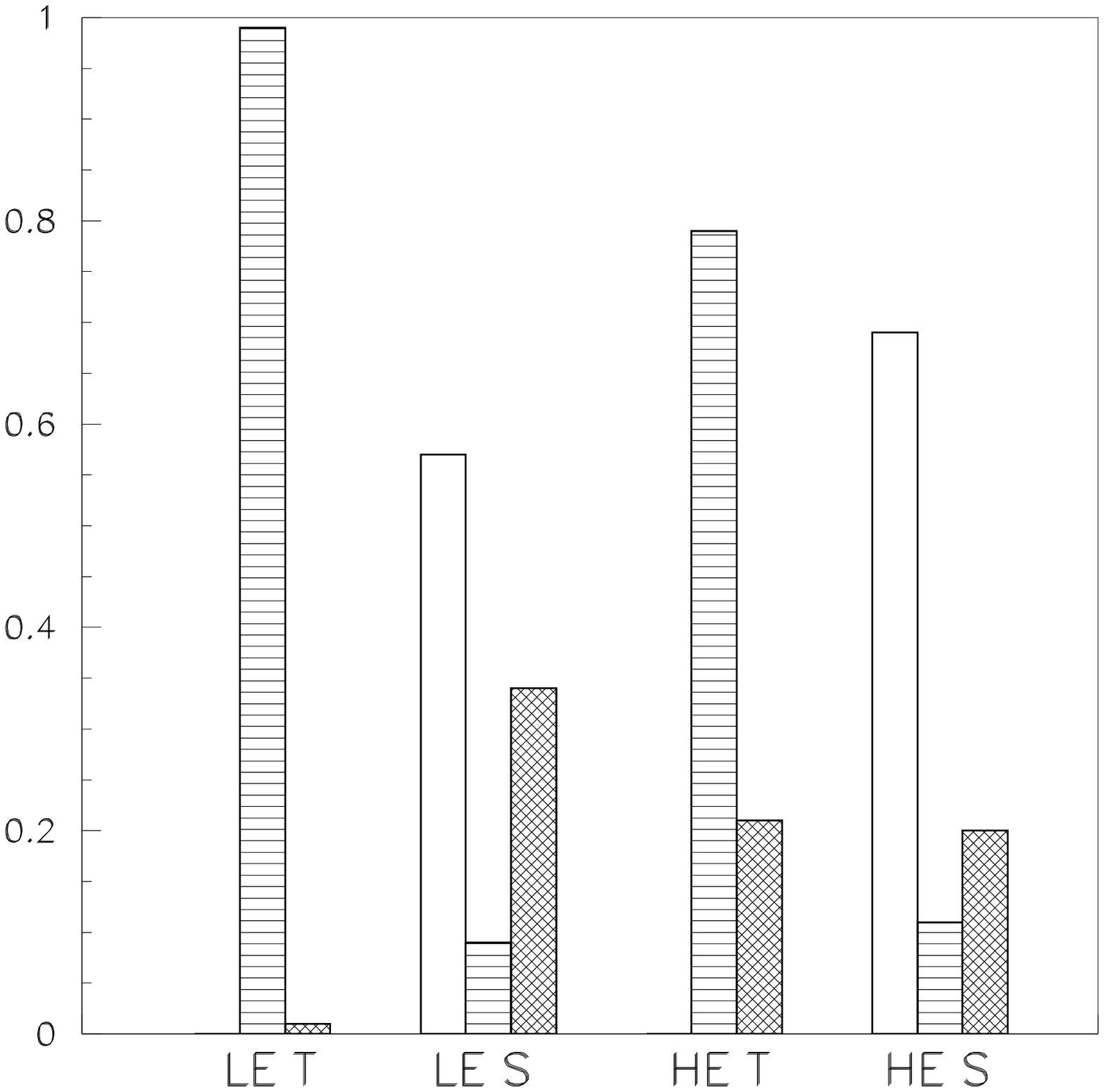,width=6cm}
\end{tabular}
\caption{In the left panel we give the zenith angle distribution
of events of different kind and at different energy for the
fiducial case $A=B=C=D=1$ and in one year of collection time. In
the right panel we report the fractional contribution of the three
flavors (empty, horizontal, and full hatching for electrons,
muons, and taus, respectively) to the different type of events  (T
stands for tracks, S for showers), assuming flavor equipartition
in the incoming fluxes.} \label{fig2}
\end{figure}

We end up with four different categories of events: LE tracks, LE
showers, HE tracks, and HE showers. In the left panel of Fig.
\ref{fig2} we illustrate the zenith angle distribution of the
events of different kind and at different energy for the fiducial
case $A=B=A'=B'=C=D=1$ and in one year of collection time. First
note that, in the case of an equal cross section in the two bins,
the corresponding LE and HE curves should have the same shape,
differing only in the normalization. The fact that this is not the
case is a manifestation of the separate sensitivity of NTs to the
cross section {\it and} the flux. The angular shapes of these
curves are plausible, too. For downgoing  events ($\cos\theta >0$)
the interaction probability of neutrinos is $\ll$1. Tracks can
propagate into the detector even when forming relatively far from
it; thus, the larger the grammage crossed the more numerous the
events. The most notable manifestation of this effect is given by
the peak near the horizon ($\cos\theta\simeq 0$) due to the so
called Earth-skimming events. The distribution of downgoing
showers is instead flat. This is because contained events are the
dominant contribution to this class, and the previous enhancement
factor due to leptons created far away is not present. Instead,
the distribution of  the upgoing events ($\cos\theta<0$) is more
and more suppressed towards $\cos\theta=-1$. This is due to the
fact that at these energies the Earth is opaque to neutrinos, an
effect most pronounced for events closer to the nadir. For tracks
the suppression is even more pronounced, since it involves not
only the suppression of the neutrino flux entering the detector,
but also the additional energy loss for charged lepton tracks
forming outside the instrumented volume. Comparing LE and HE
events, note that  HE events suffer a higher opacity, so the
upgoing rate suppression is more pronounced with respect to the LE
case. On the other hand, the range of charged leptons is larger,
which explains the steeper behavior of the HE downgoing tracks
curve.  The number of downgoing track events also grows at HE,
since the cross section is higher and we are in a regime where the
interaction probability is $\ll 1$. The spiky bumps at
$\cos\theta\simeq \pm 0.6$ are simply an effect of the cubic
geometry assumed (there is more instrumented volume when looking
at angles closer to a vertex of the cube). In the right panel of
Fig.~\ref{fig2}, we show the flavor composition of the four
classes of events. As expected, basically only muons contribute to
the LE track bin. Indeed, the tau decay length is $D_\tau\simeq
50\,{\rm m}(E_\tau/{\rm PeV})$. A realistic cut due to the spacing
of the towers would probably remove even this small tau
contamination. On the other side, the tau contribution to shower
events decreases with energy, due to the increase of the
relativistic tau gamma factor. Although we shall not discuss this
point further here, it has been noted that the fact that different
types of events are not flavor blind offers another possibility
for astrophysical diagnostics \cite{flavors}.

Although the overall number of events at a NT is limited, the
previous discussion suggests that at very least a further
partition of each of the previous  bins in upgoing or downgoing
events would greatly improve the diagnostic power for
disentangling cross section from flux. Note that it is indeed a
very weak requirement for a NT to be able to reconstruct at least
the sign of the cosine of the zenith angle, in order to reject the
background from downgoing muons from cosmic ray showers. Even so,
it is at present unclear how well one can tag {\it downgoing}
events as a signal against the cosmic-ray background. Around PeV
energies, this should be feasible in IceCube, where the IceTop
surface array offers a veto for cosmic ray events. For a
Mediterranean km$^3$ neutrino telescope no final design exists,
but the possibility has been suggested to extend the detection
capability of the experiment with some sea top stations
\cite{SeaTop}. Keeping this possibility in mind, we shall consider
both the case in which downgoing signal events can be used and the
one where this is not possible. Also, we have tested that slightly
changing the angular intervals and/or adding a third angular bin
have only a minor effect on the sensitivity in the parameter
space.

For each choice of parameters, we end up with the counts
$N^{K}_{i,\alpha}$, where $K$ labels the type/topology of the
event, $i$ the energy bin and $\alpha$ the angular one. As already
mentioned, we always consider two energy bins: a low-energy one,
corresponding to $E_1\leq \Delta E \leq E_2$ and a high-energy
one, corresponding to $\Delta E >E_2$. Regarding the choice of the
angular binning, when both the downgoing and upgoing information
is used, we divide the angular range in the two bins
$[0^\circ,90^\circ]$ and $[90^\circ,180^\circ]$ while, if only
upgoing events are included in the analysis, we consider the two
bins $[90^\circ,107^\circ]$ and $[107^\circ,180^\circ]$. We fix
$C=1$ since it is just a normalization, simply correlated to the
exposure time needed to achieve the proper event statistics. The
remaining parameters vary in a grid of 125+125 theoretical models
as follows: $D\in [0.5,1.5]$ in steps of 0.25, while $A=A', B=B'
\in[0,4]$ in steps of 1 or $A=B=1,A',B' \in[0,4]$ in steps of 1 in
the two cases under investigation, that is: 1) CC and NC cross
sections vary proportionally to each other; 2) only NC cross
section changes. To determine approximately the range of
parameters, we require that at least one event falls in each bin
in one year of running. We then compare the counts
$N^{K}_{i,\alpha}$ with the expected counts $C^{K}_{i,\alpha}$ for
the benchmark case $A=B=A'=B'=C=D=1$ by means of a multi-Poisson
likelihood analysis \cite{Baker}, in which the likelihood
function, ${\cal L}\propto\exp(-\chi^2/2)$, is defined using the
following expression for the $\chi^2$:
\be \chi^2=2\sum_{i,\alpha,K}\left[
N^{K}_{i,\alpha}-C^{K}_{i,\alpha}+C^{K}_{i,\alpha}\ln\left(
\frac{C^{K}_{i,\alpha}}{N^{K}_{i,\alpha}}\right)\right].
\ee
%

%%%%%%%%%%%%%%%%%%%%%%%%%%%%%%%%
\section{Results}\label{results}
%%%%%%%%%%%%%%%%%%%%%%%%%%%%%%%%

\begin{figure}[!t]
\begin{tabular}{ccc}
\epsfig{file=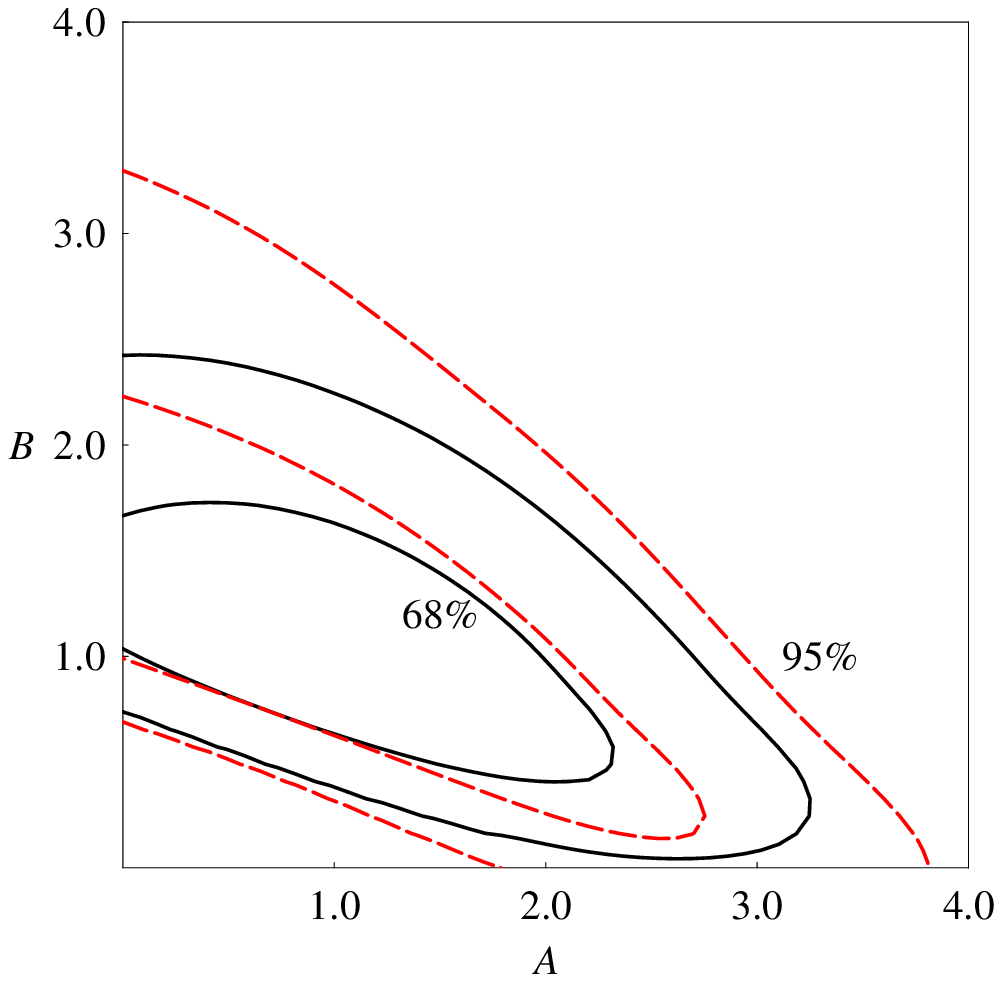,width=5cm} &
\epsfig{file=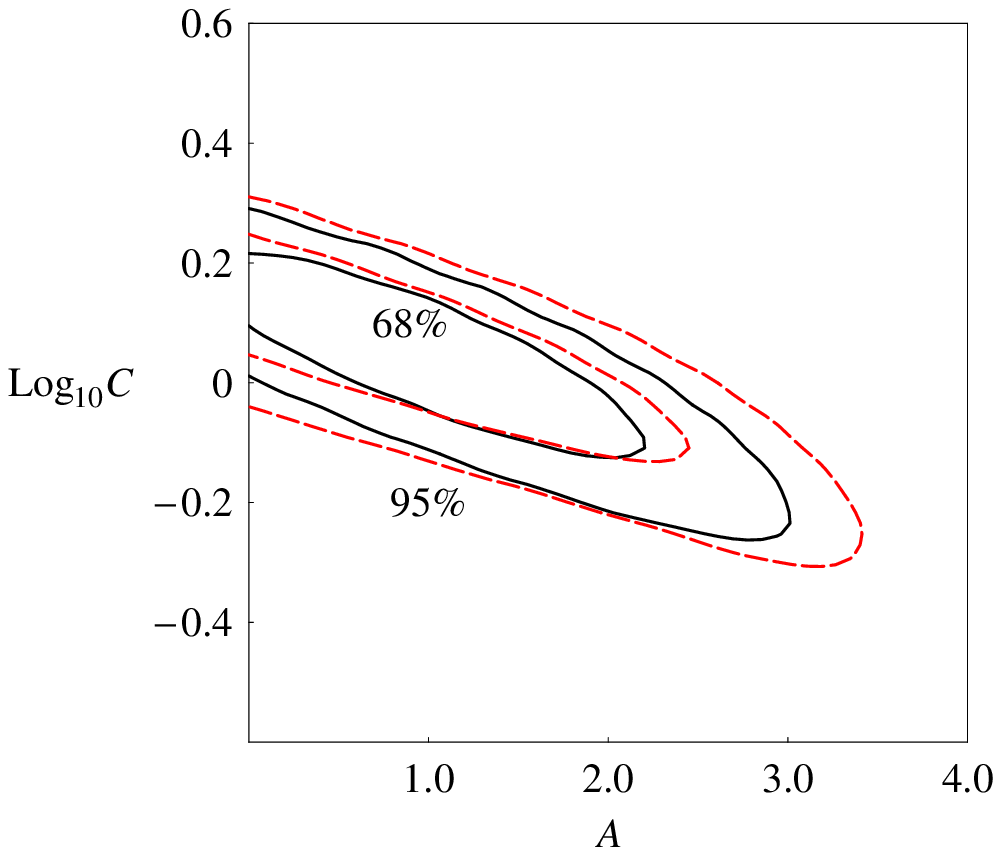,width=5cm} &
\epsfig{file=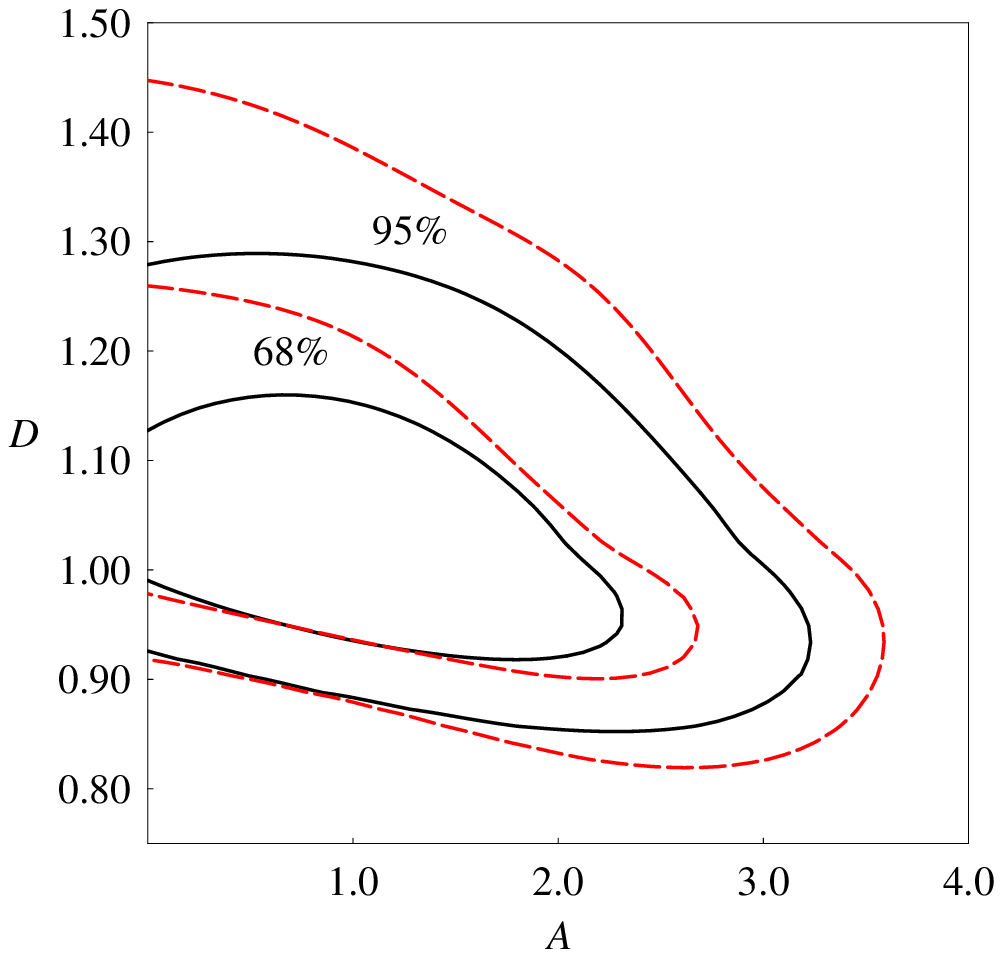,width=5cm} \\
\epsfig{file=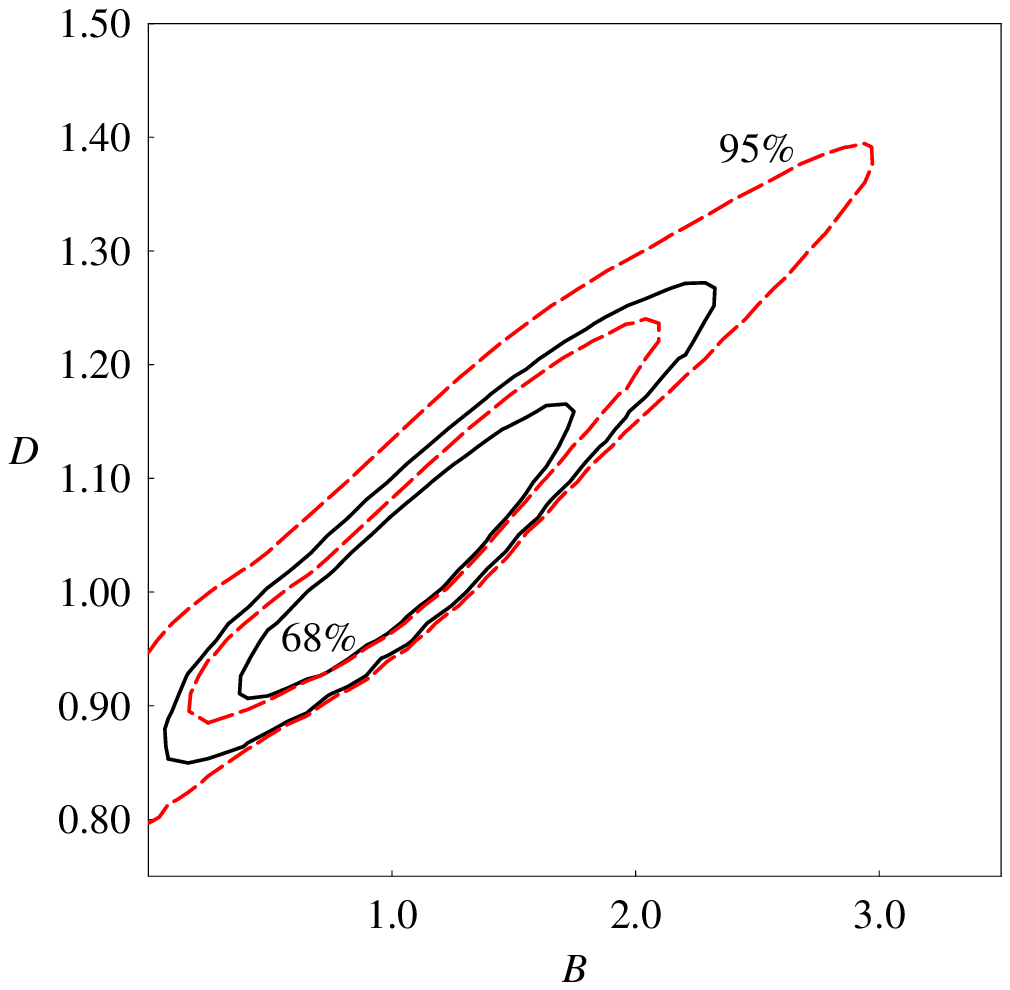,width=5cm} &
\epsfig{file=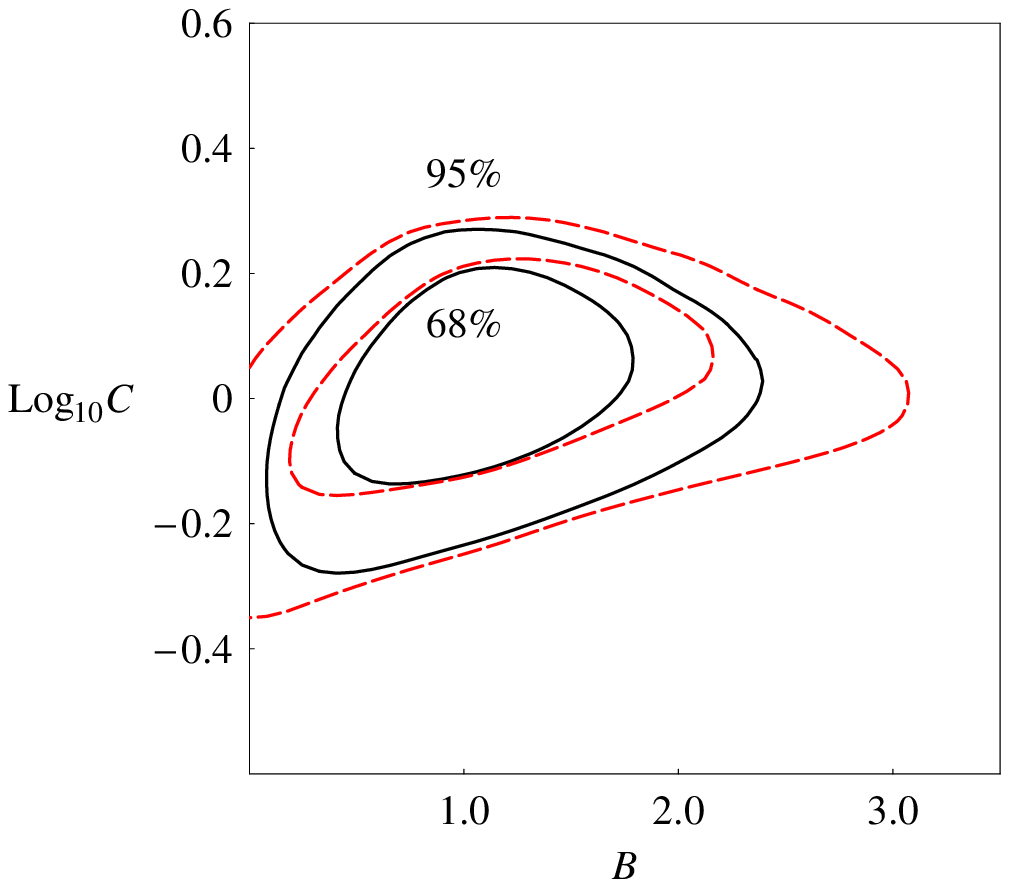,width=5cm} &
\epsfig{file=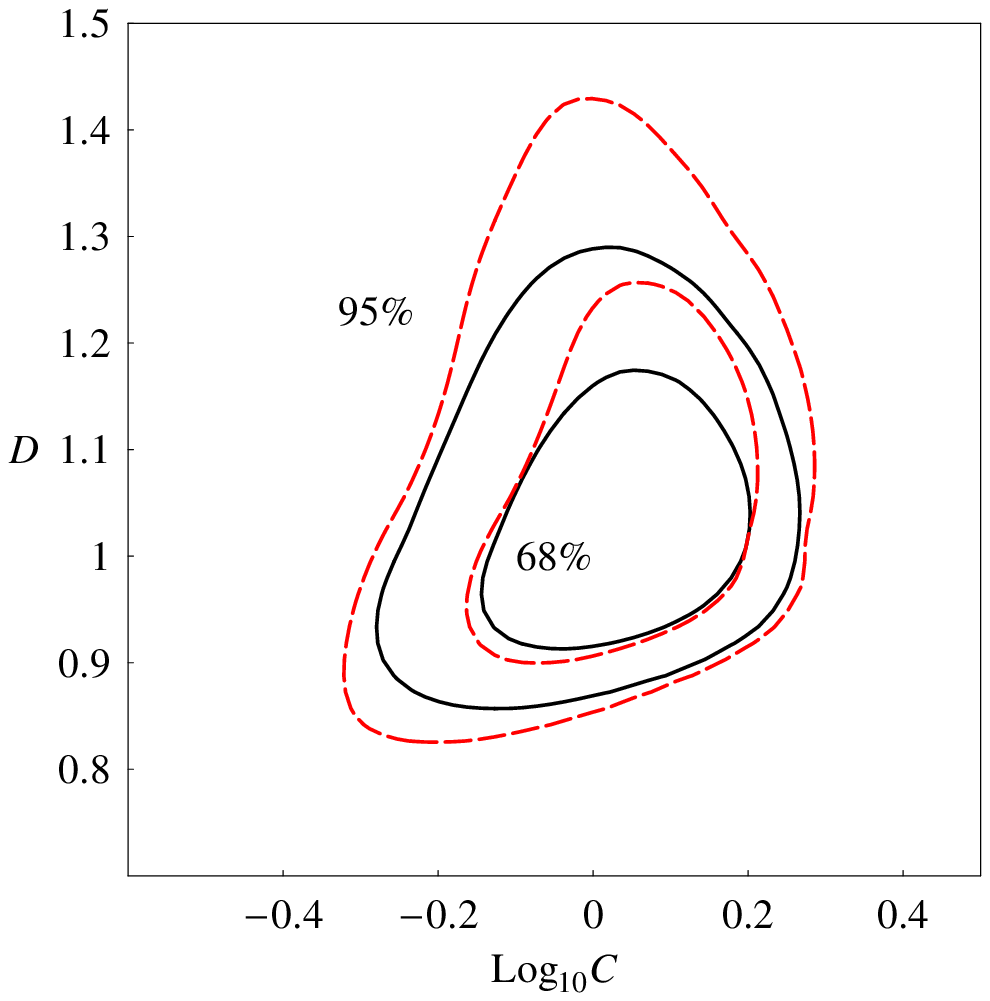,width=5cm}
\end{tabular}
\caption{For the case where CC and NC cross sections change
proportionally to each other, the solid (black) contours (68\% and
95\% CL) represent marginalized constraints on pairs of parameters
$\{A,B,C,D\}$ in the case where both upgoing and downgoing
neutrino events can be tagged, for five years of exposure. Dashed
(red) contours represent the case where no topological information
is assumed, i.e. when one sums track and shower events.}
\label{contour1cc}
\end{figure}

\begin{figure}[!t]
\begin{tabular}{ccc}
\epsfig{file=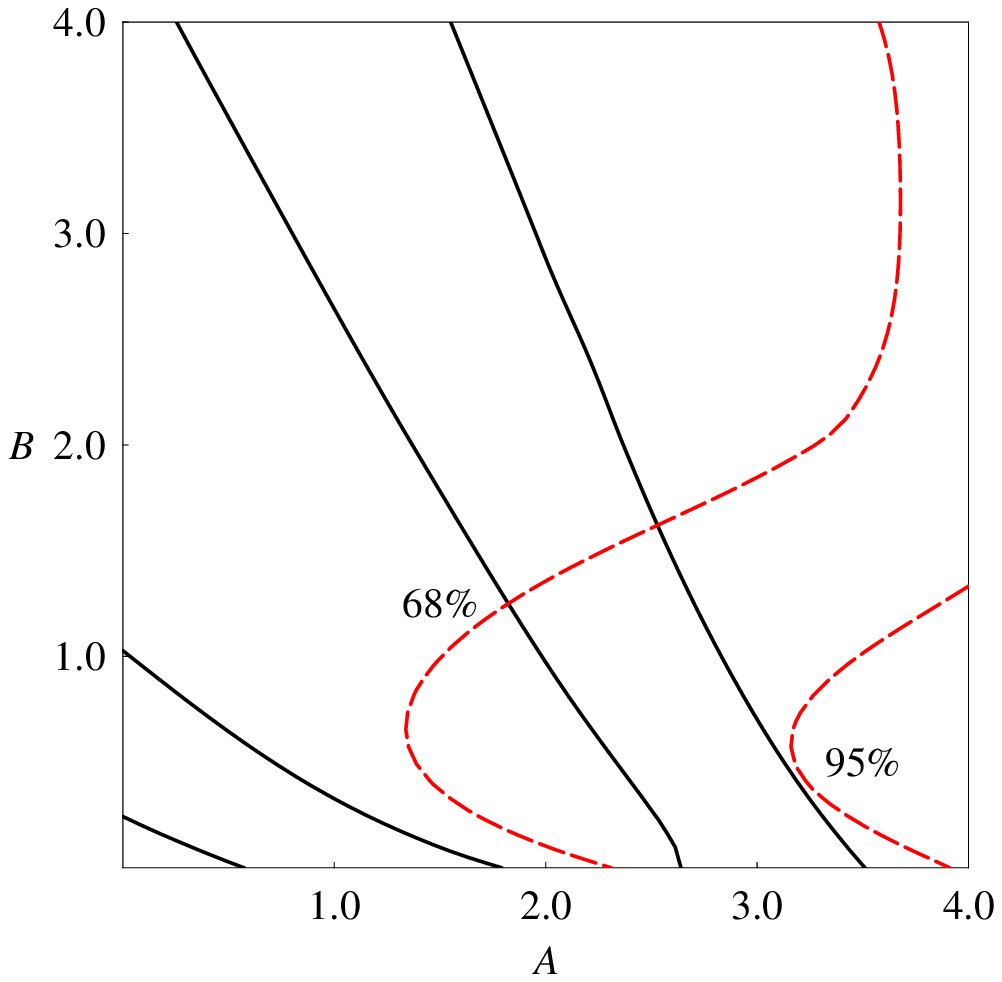,width=5cm} &
\epsfig{file=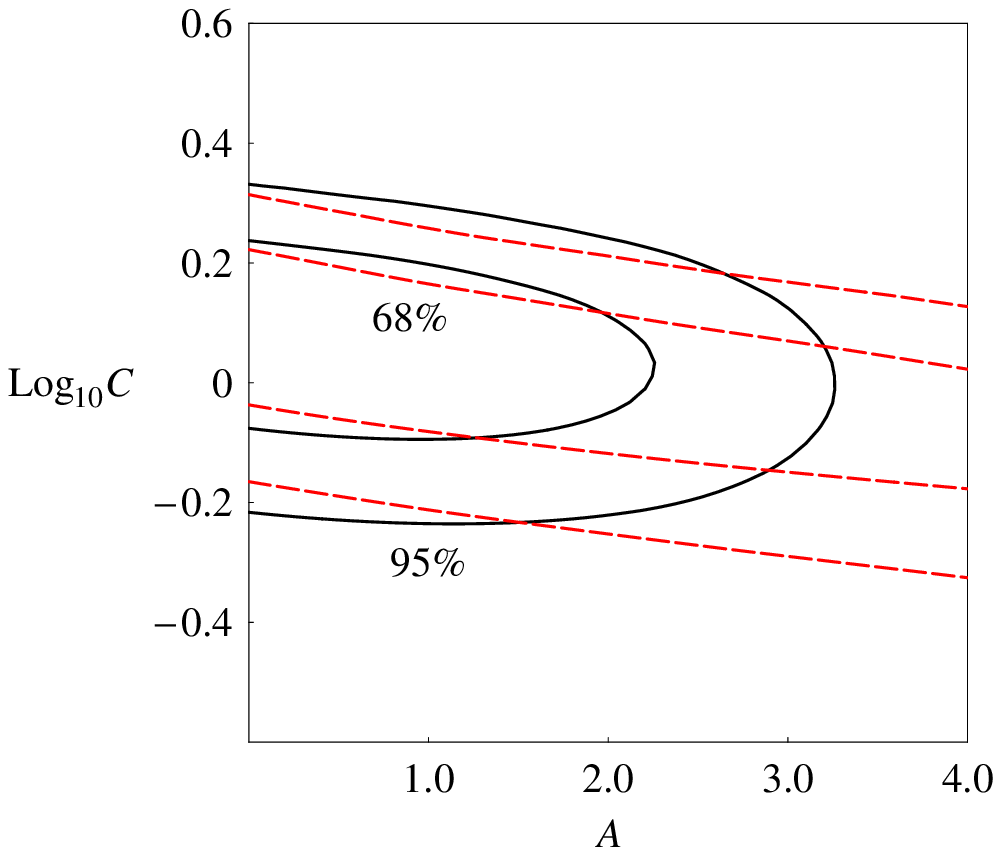,width=5cm} &
\epsfig{file=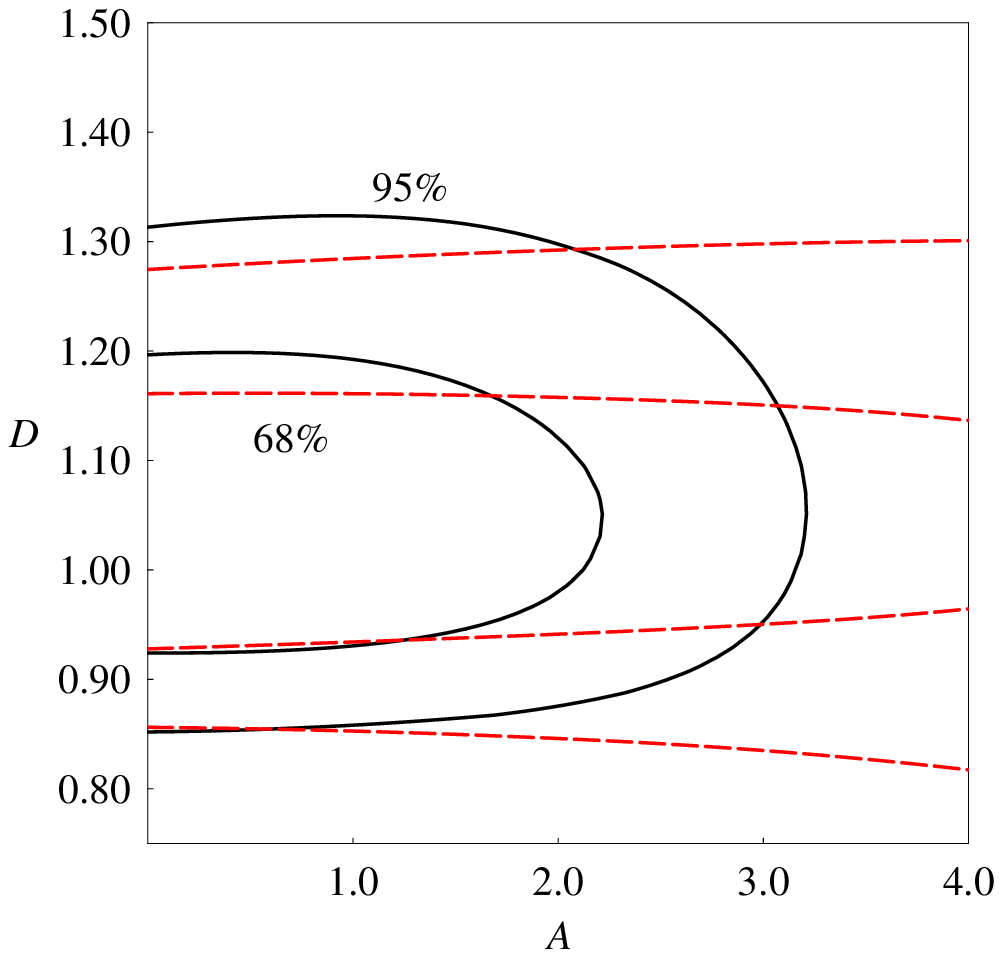,width=5cm} \\
\epsfig{file=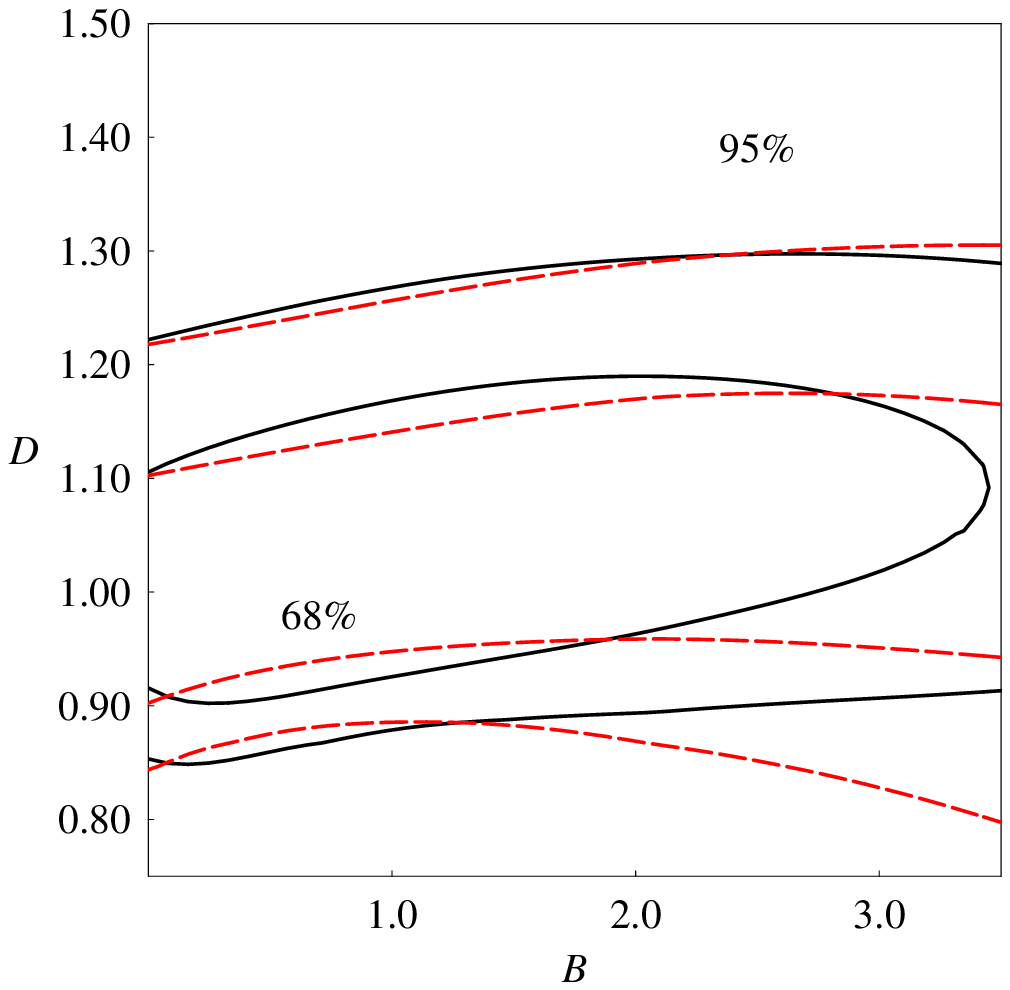,width=5cm} &
\epsfig{file=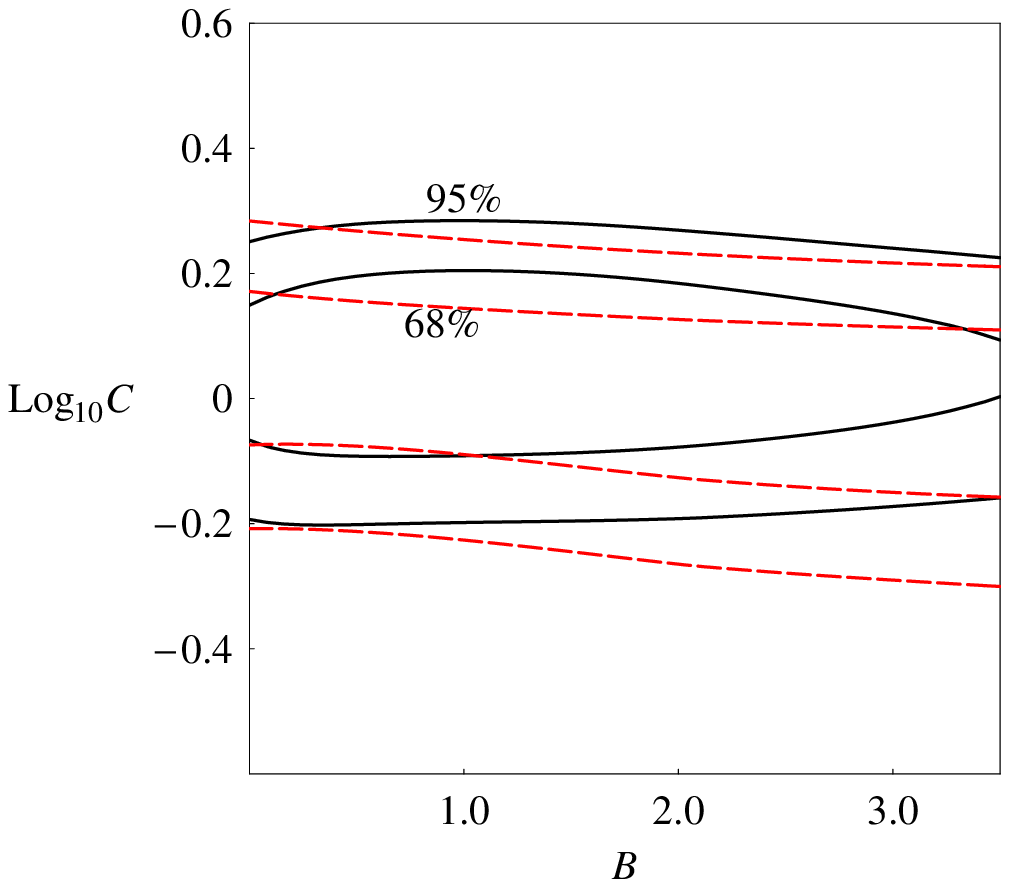,width=5cm} &
\epsfig{file=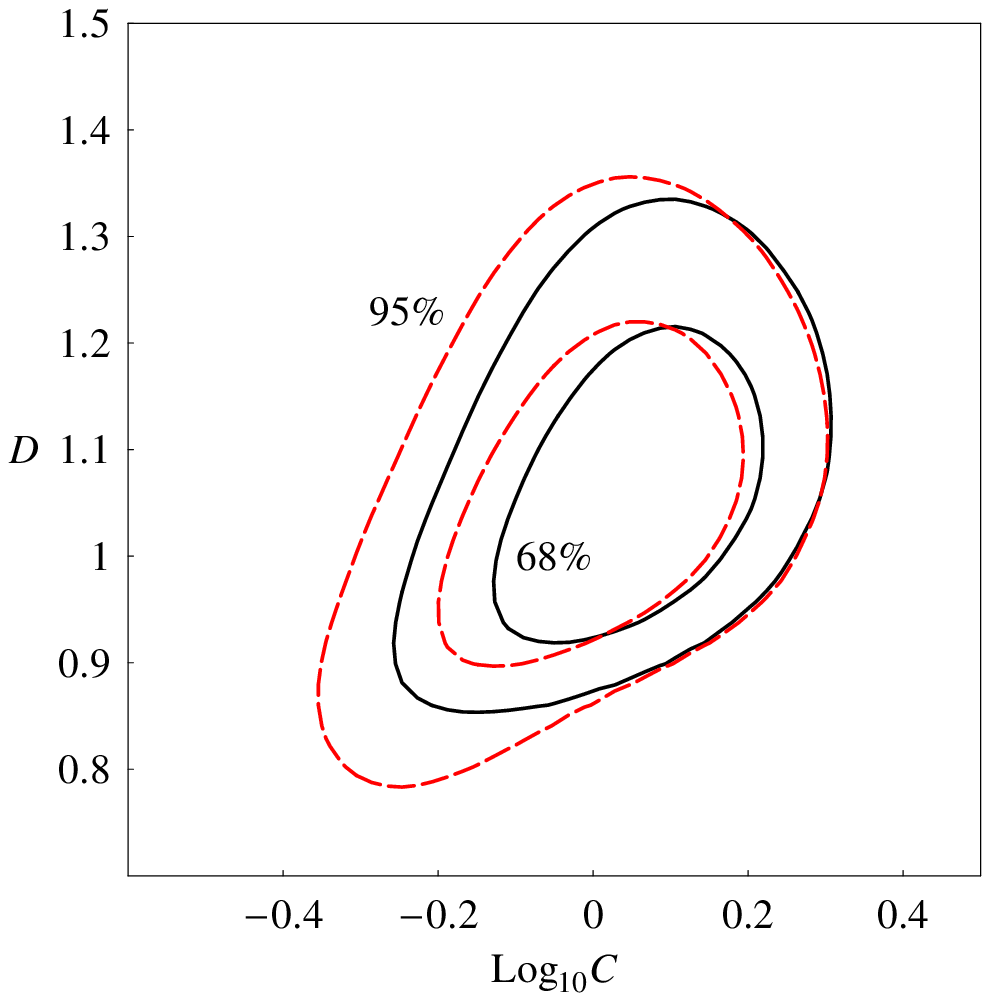,width=5cm}
\end{tabular}
\caption{For the case where CC and NC cross section change
proportionally to each other, we report the marginalized
constraints on couples of parameters $\{A,B,C,D\}$ (contours at
68\% and 95\% CL) in the case where only upgoing neutrino events
can be tagged and split in two angular bins, for five years of
exposure. Track vs. shower separation capability is assumed.
Dashed (red) contours represent the previous case where all events
are summed in a single angular bin.} \label{contour2cc}
\end{figure}

\begin{table}[b]
\begin{tabular}{ccc}
\begin{tabular}{ccccc}
\hline\hline\\[-0.2cm]
& $\qquad$                                         &   68\% CL &   $\qquad$    &   95\% CL     \\[0.2cm]   \hline\hline\\[-0.2cm]
$A$            & &  $1.0^{+0.6}_{-0.7}$           & & $1.0^{+1.2}_{-1.0}$              \\[0.2cm]   \hline\\[-0.2cm]
$B$                      & &    $1.2^{+0.4}_{-0.5}$  & & $1.2^{+0.7}_{-0.8}$   \\[0.2cm]   \hline\\[-0.2cm]
$\log_{10}{C}$ & &  $0.04 \pm 0.10$ & & $0.04^{+0.15}_{-0.19}$   \\[0.2cm] \hline\\[-0.2cm]
$D$ &                    & $1.0\pm 0.1$     & & $1.0^{+0.2}_{-0.1}$ \\[0.2cm]   \hline\hline\\
\end{tabular}
& $~~~~~~~~~~~~~~~$ &
\begin{tabular}{ccccc}
\hline\hline\\[-0.2cm]
& $\qquad$ & 68\% CL    &   $\qquad$    &   95\% CL  \\[0.2cm]  \hline\hline\\[-0.2cm]
$A'$ & & $ < 2.0$                    & & $< 3.4$             \\[0.2cm] \hline\\[-0.2cm]
$B'$ &                    & $<1.4$      & & $< 2.4$     \\[0.2cm] \hline\\[-0.2cm]
$\log_{10}{C}$ & &  $-0.01^{+0.11}_{-0.13}$ & & $-0.01^{+0.20}_{-0.24}$ \\[0.2cm] \hline\\[-0.2cm]
$D$ &                    & $1.1\pm 0.1$    & & $1.1\pm{0.2}$  \\[0.2cm] \hline\hline\\
\end{tabular}
\end{tabular} \caption{Determination of cross section and flux parameters,
obtained by marginalizing the (black) solid contours of Fig.
\ref{contour1cc} (left table) and of Fig. \ref{contour1nc} (right
table).}\label{table1}
\end{table}

In this Section we present the results of our analysis. Fig.s
\ref{contour1cc} and \ref{contour2cc} are obtained considering the
case where CC and NC cross sections change proportionally to each
other. The solid (black) curves in the panels of Fig.
\ref{contour1cc} show the marginalized constraints on couples of
parameters (contours at 68\% and 95\% CL) in the case one has both
upgoing and downgoing information available, for five years of
exposure. Dashed (red) contours represent the case where no
topological information is assumed, i.e. when one sums track and
shower events. In general, there is a negative correlation between
the cross section parameters $A$ and $B$, since to some extent a
higher cross section in the range $E_1<E<E_2$ can be compensated
by a smaller one in the higher energy range (note also that in the second line
of Eq.~(\ref{CC}) both $A$ and $B$ enter). Indeed, the primary
energies are not directly observable , while the energy losses $\Delta E$
are.  Also, there are negative correlations between $A$ and $C$ or
$D$. The reason is that a higher cross section
bin can be somewhat compensated by a smaller flux normalization
($C$) or a slightly steeper flux spectrum ($D$). The
anti-correlation between $A$ and $B$ explains the correlations
between $B$ and $C$ and $B$ and $D$. In all cases the contours
close, and the parameters can be determined to some accuracy. The
degeneracy between flux and cross section is thus broken by the
combined use of angular and energy independent information. Note
how the topological information adds diagnostic power,
although does not change the qualitative behavior.

\begin{figure}[!htp]
\begin{tabular}{ccc}
\epsfig{file=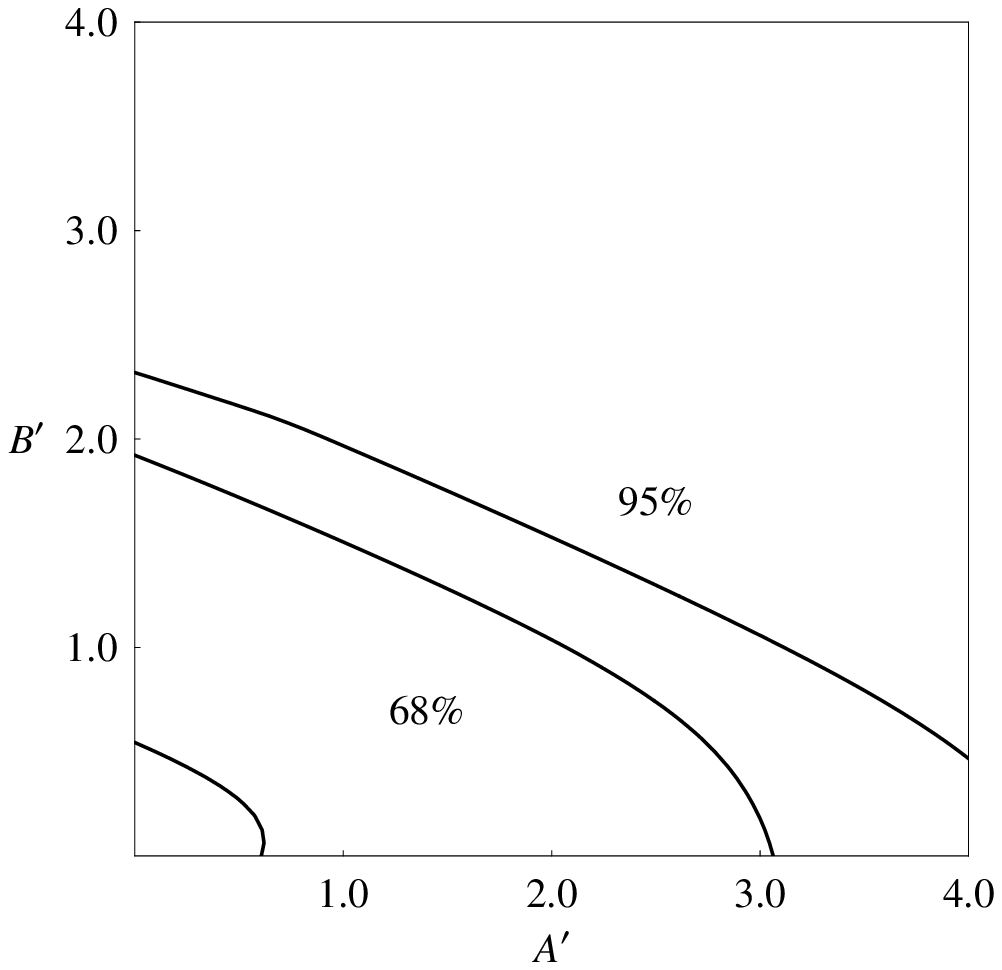,width=5cm} &
\epsfig{file=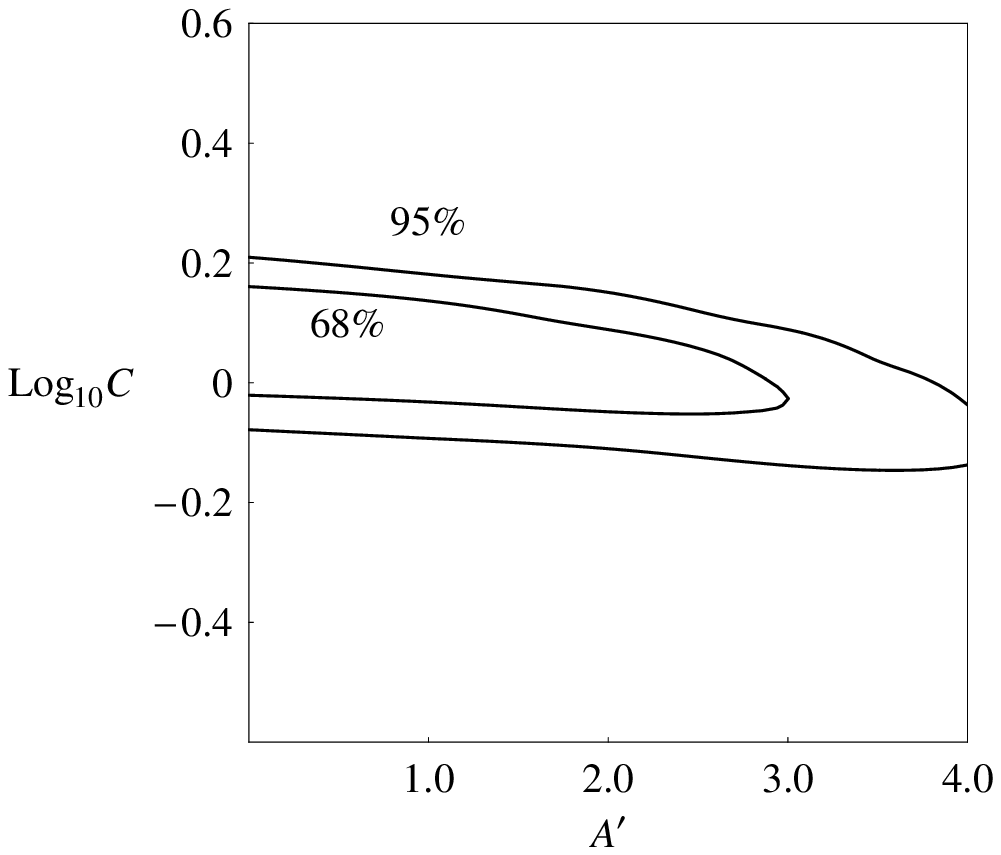,width=5cm} &
\epsfig{file=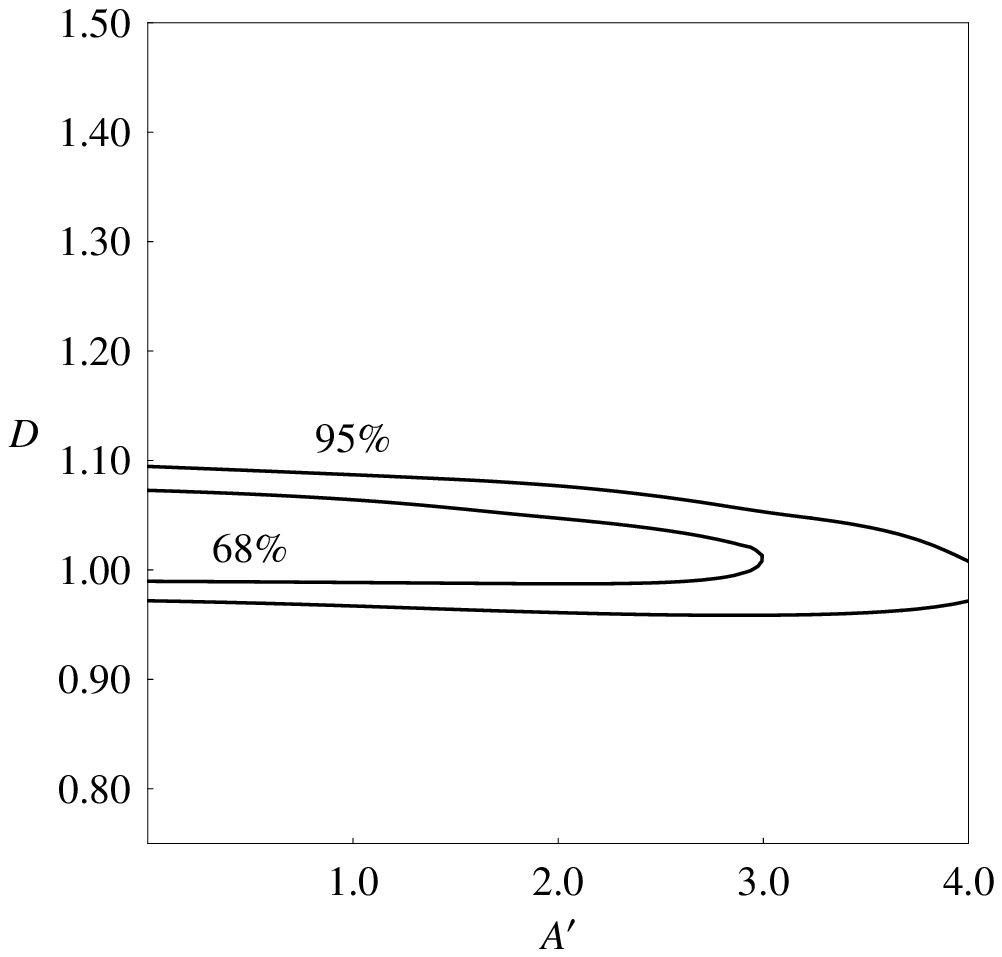,width=5cm} \\
\epsfig{file=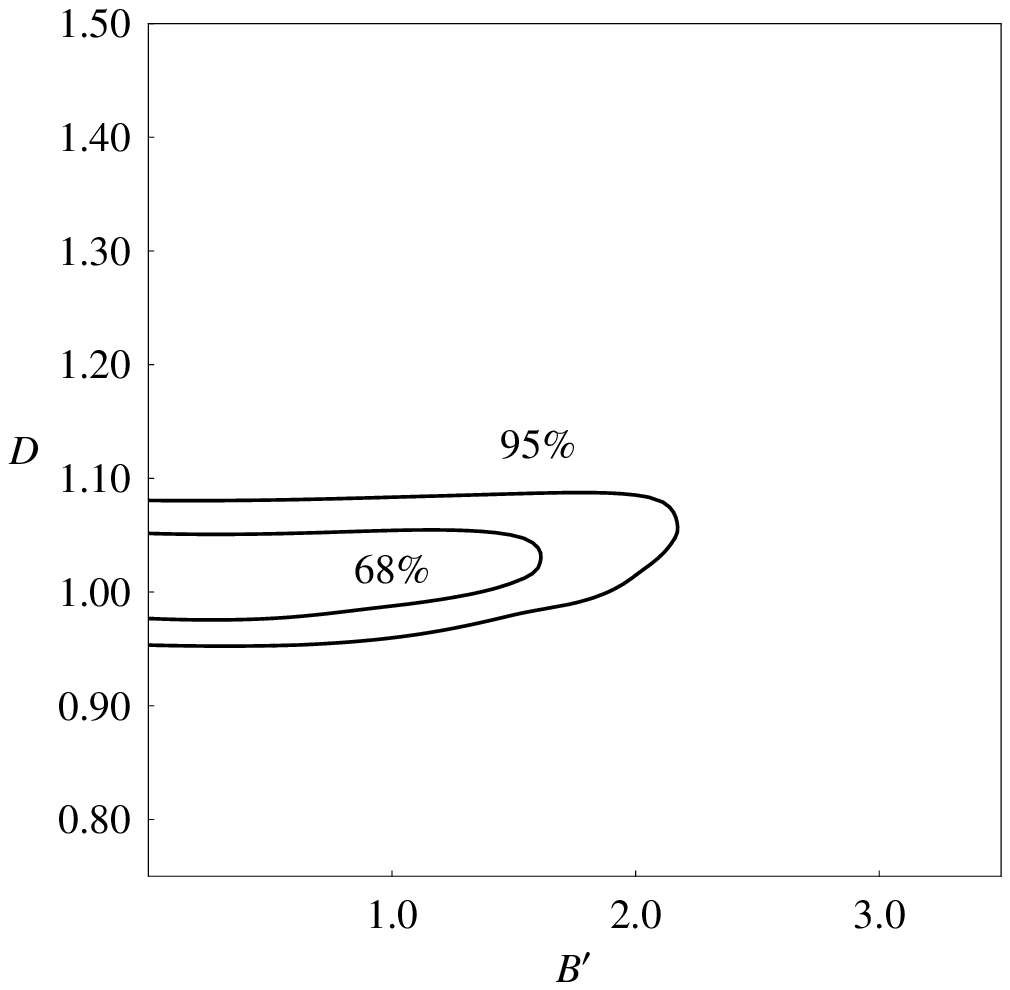,width=5cm} &
\epsfig{file=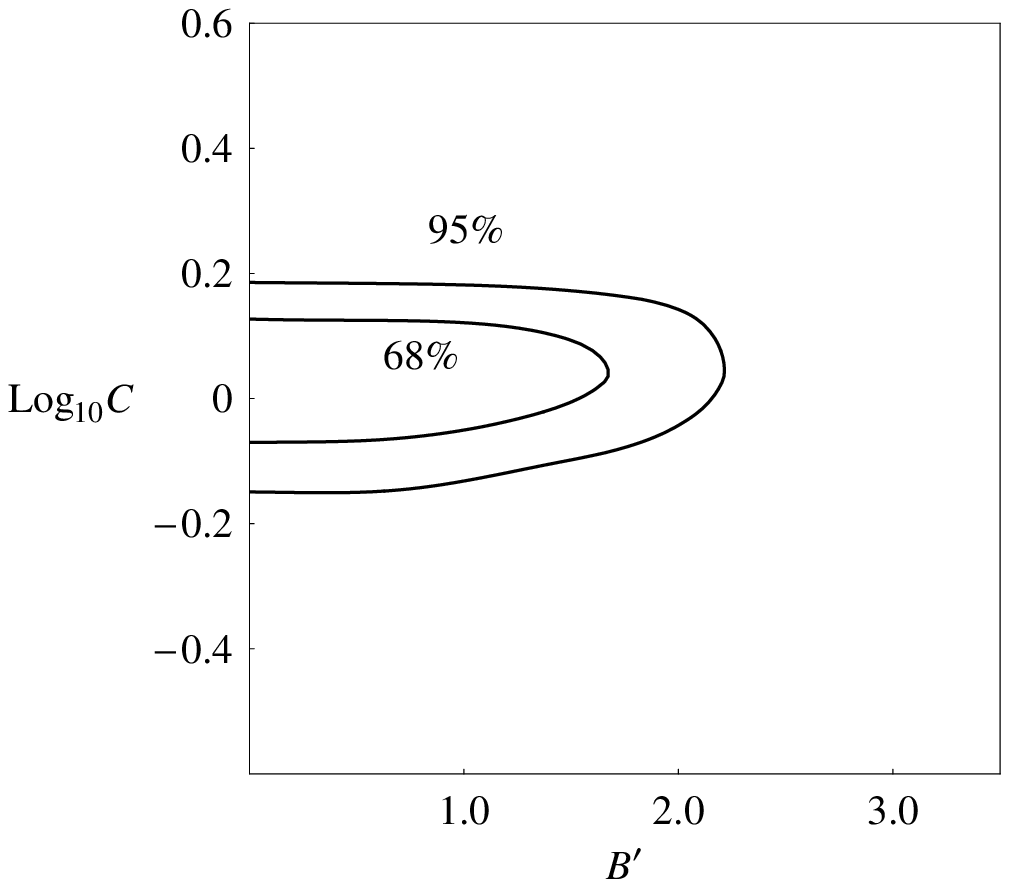,width=5cm} &
\epsfig{file=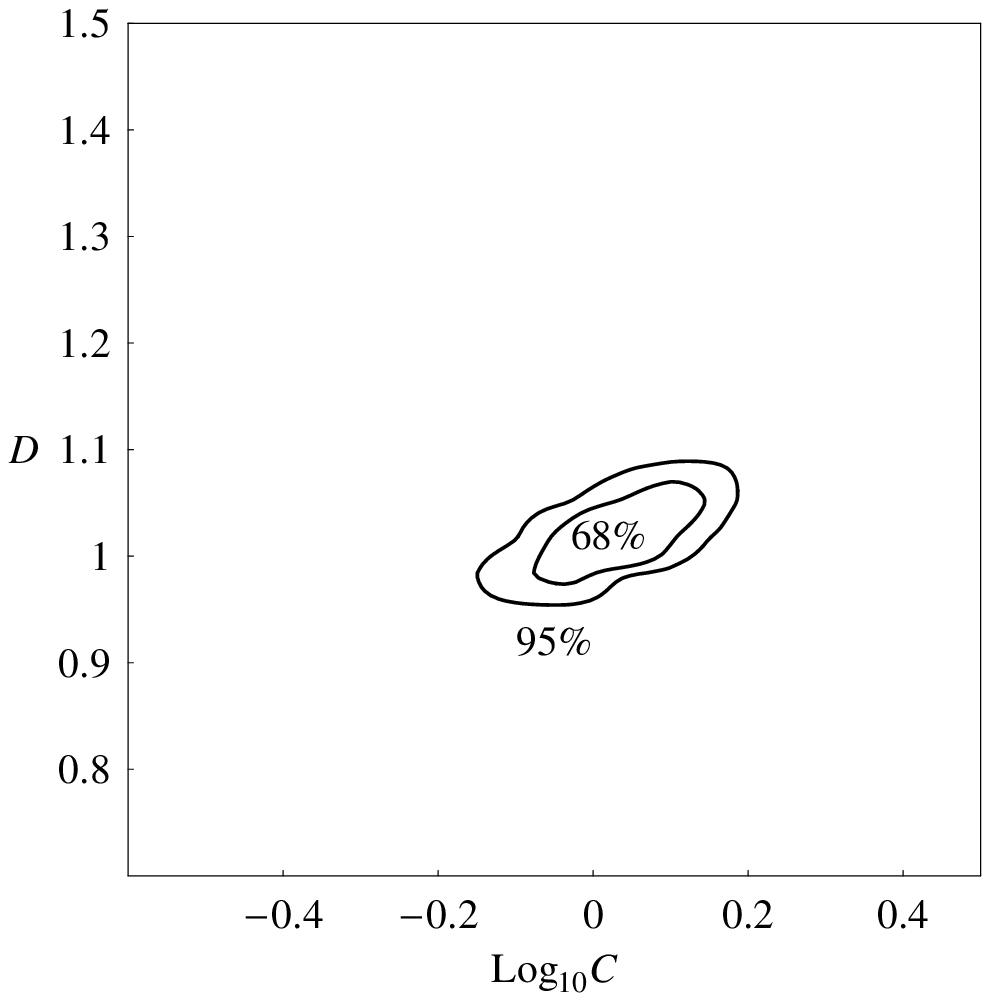,width=5cm}
\end{tabular}
\caption{For the case where only NC cross section changes while CC
one assumes its standard value, we show the marginalized
constraints on couples of parameters $\{A',B',C,D\}$ (contours at
68\% and 95\% CL) in the case where  both upgoing and downgoing
neutrino events can be tagged, for five years of exposure.}
\label{contour1nc}
\end{figure}

\begin{figure}[!htp]
\begin{tabular}{ccc}
\epsfig{file=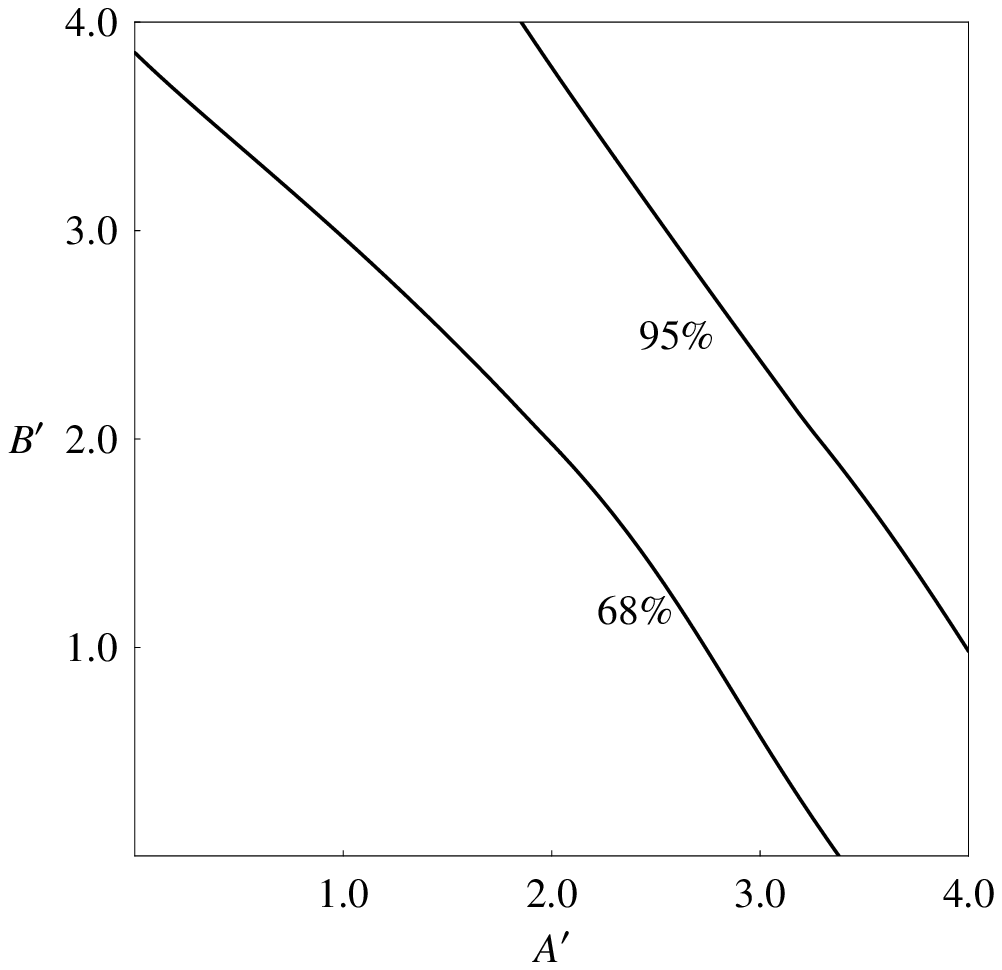,width=5cm} &
\epsfig{file=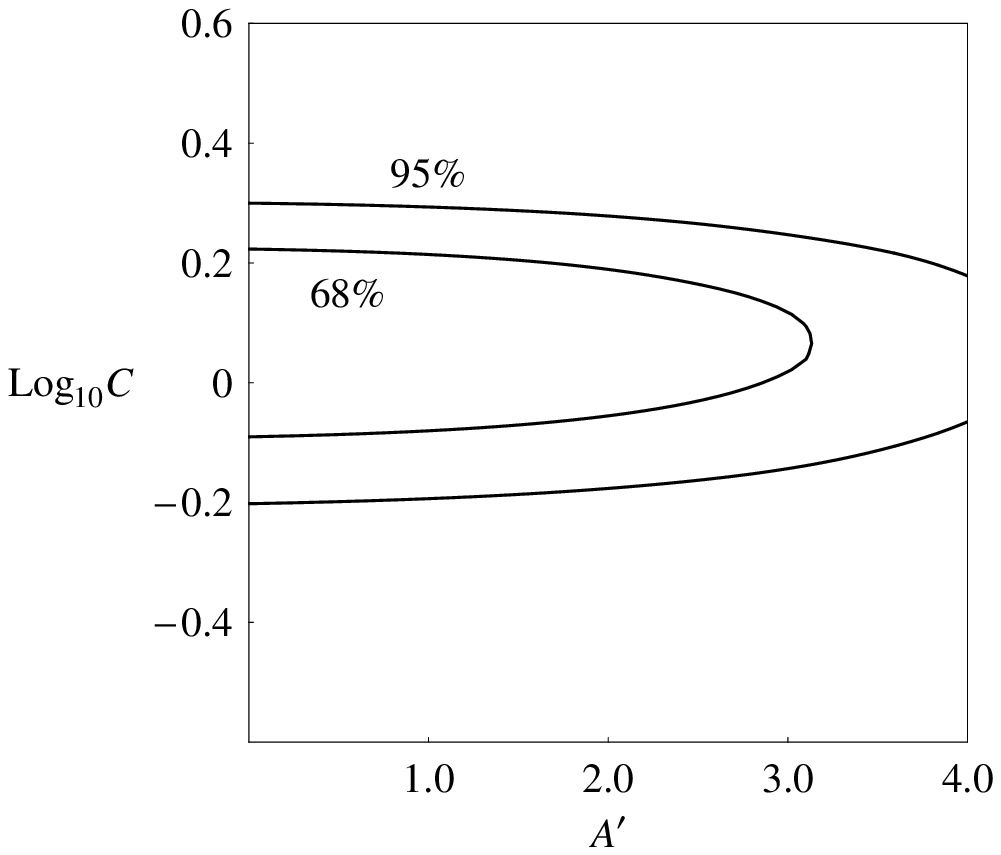,width=5cm} &
\epsfig{file=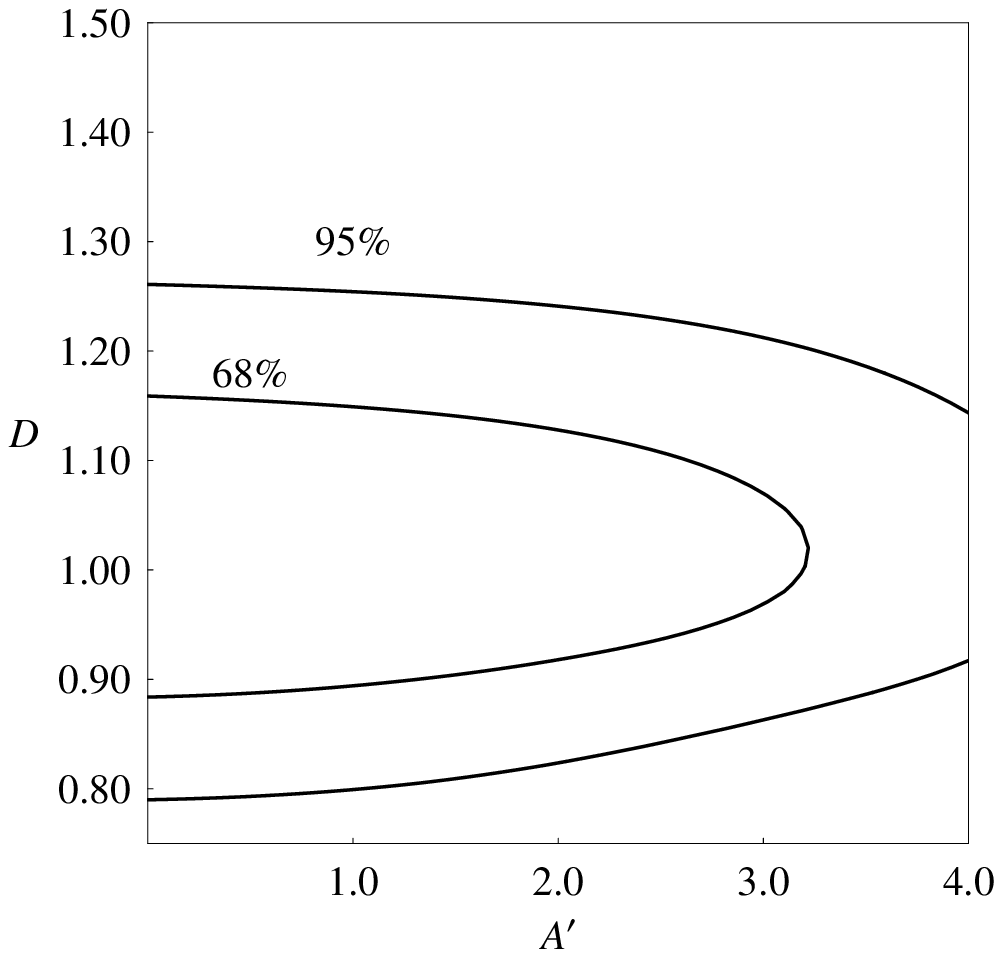,width=5cm} \\
\epsfig{file=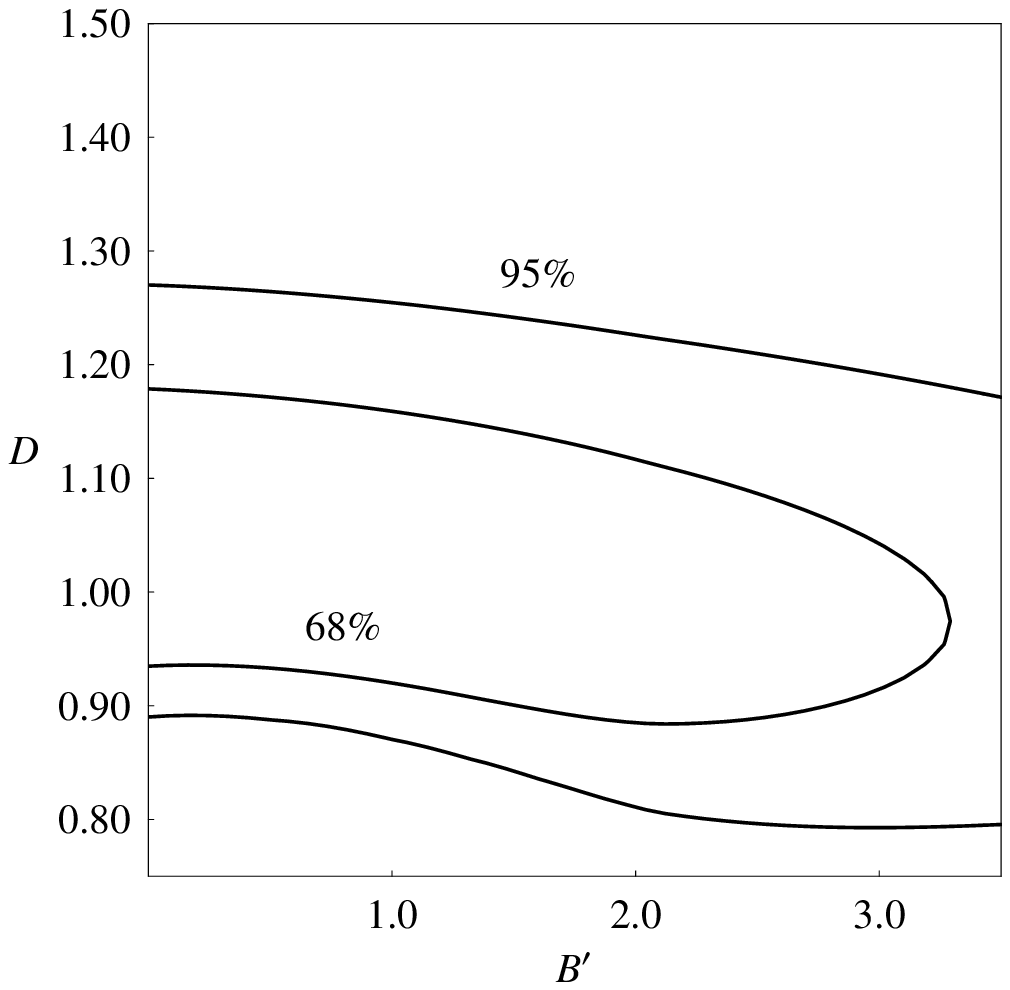,width=5cm} &
\epsfig{file=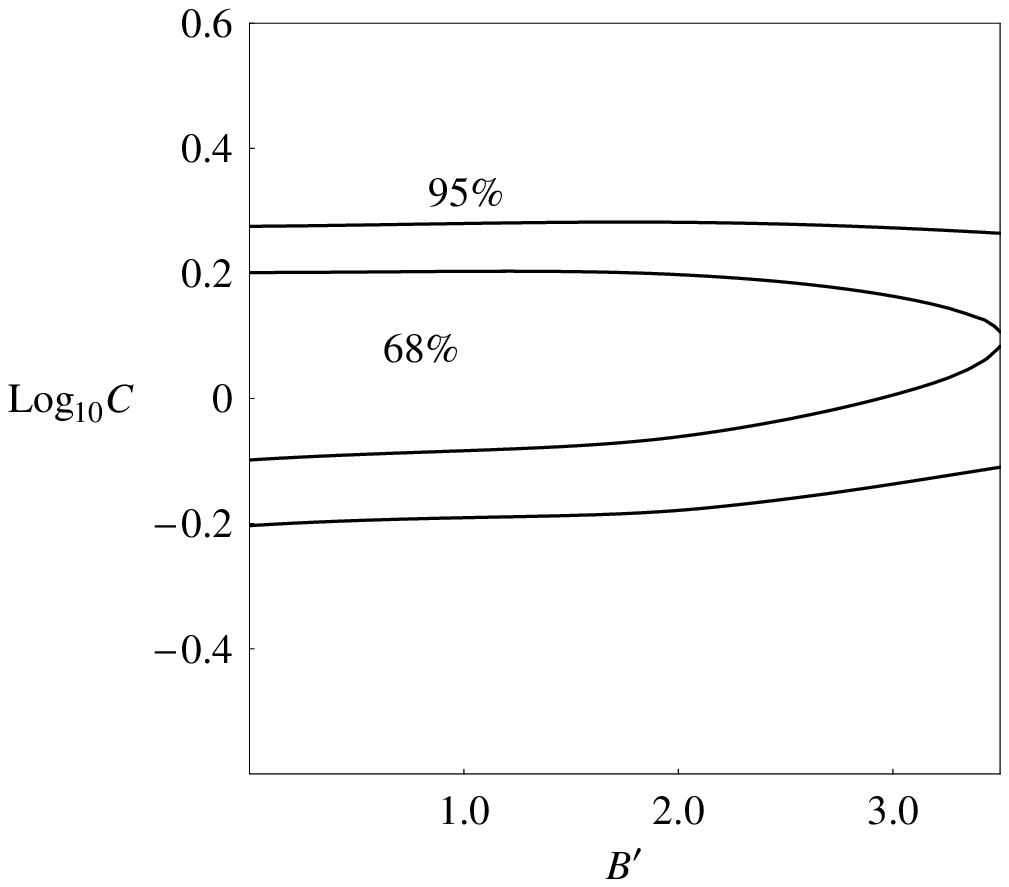,width=5cm} &
\epsfig{file=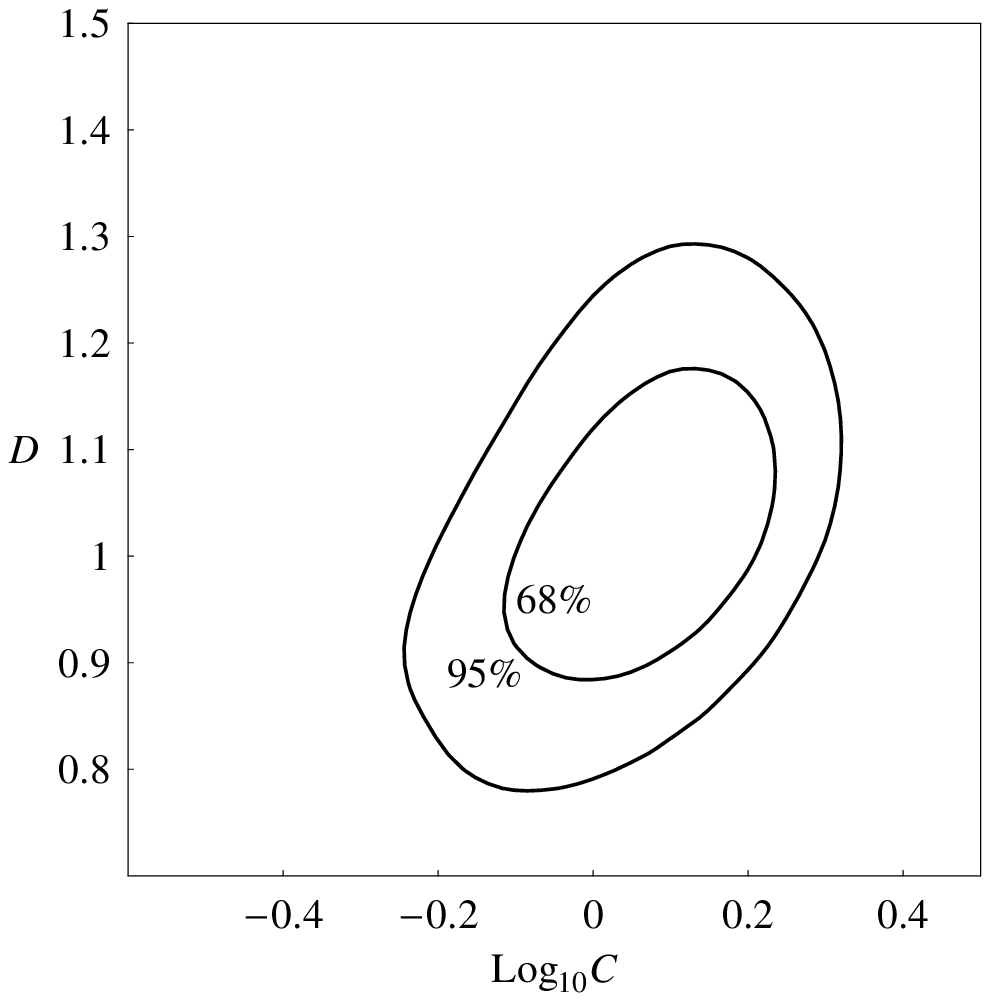,width=5cm}
\end{tabular}
\caption{Same of Fig. \ref{contour1nc} but in the case where only
upgoing neutrino events can be tagged and split in two angular
bins, for five years of exposure. Track vs. shower separation
capability is assumed.} \label{contour2nc}
\end{figure}

In Fig. \ref{contour2cc} we show instead the marginalized
constraints on couples of parameters (contours at 68\% and 95\%
CL) in the case where only upgoing events can be tagged, again for
five years of exposure. Solid (black) curves correspond to the
case of two bins, $[90^\circ,107^\circ]$ and
$[107^\circ,180^\circ]$, while dashed (red) curves show the
results when these two bins are merged together. Only the flux
parameters can be determined to some accuracy, even if some
sensitivity to $A$ and $B$ is still recovered in the case of two
angular bins. The basic physical reason for this behavior has been
discussed analytically in \cite{Hussain:2006wg}. As long as NC and
CC cross sections change proportionally, then both the upward
track rate and the upward shower rate are independent of cross
sections, with or without new physics involved.  If the energy
dependence of the cross sections involved is the same, as we
assume here for the fiducial case, the two observables (upgoing
track and shower events) are just proportional to the integrated
flux in the corresponding energy bin.

The point we want to emphasize here is that without the tagging
capability of downgoing events, there is little hope for NTs to
get meaningful constraints on the cross sections. A further
binning of the upgoing events into two angular bins may only help
a little bit, with the additional problem that the statistics is
more limited in each bin.

Finally, the curves in Fig.s \ref{contour1nc} and \ref{contour2nc}
correspond to the solid (black) lines in Fig.s \ref{contour1cc}
and \ref{contour2cc}, but assuming that only NC are changed (for
illustrative purposes, we fix here $A=B=1$ for CC). The flux
parameters $C$ and $D$ are quite accurately determined, in
particular $D$ is fixed at 10 $\%$ level, since the CC cross
section contributing to tracks and, partially, to shower events is
assumed to be known in this case. There is instead quite a poor
determination of the NC cross section parameters, and in particular of $A'$.
Indeed, only values of $A' \geq 3.4$ and $B'\geq 2.4$ are ruled out at 95\% CL.
This result is easy to understand. In fact, the largest contribution to NC
shower events corresponds to downgoing neutrinos; so the event
rate is almost independent of the intervening matter in the
neighborhood of the NT site, being simply proportional to the product of the flux
times cross section. For example, at high energies the number of NC events increases more than logarithmically as
function of the highest possible neutrino energy:  for $ B' >  D
/0.492 \sim 2 $ it becomes unacceptably large and thus excluded (actually the result
would diverge if $B'\geq 2$ in the unphysical case of an incoming
flux with $D=1$ for arbitrary high energies.)
The other relevant
feature is the sensitivity only to a combination of $A'$ and $B'$ (roughly $A'+\kappa B'$,
with $\kappa \simeq 1.5$ for the case of Fig.~5). The reason is that, due to the smaller energy transferred
in visible channels in NC events, the NC events collected in both our energy bins are mostly  sensitive
to the high energy behavior of the cross section, i.e.  the second line in Eq.~(3).
This function depends on a quasi-degenerate combination of the two parameters, explaining
the results. Of course, a larger statistics at higher energies would eventually break this degeneracy,
but would probably require too long exposure times.

In Table~\ref{table1} we summarize our results by reporting the parameter
sensitivity at 68\% and 95\% after the marginalization of the
likelihood function for both scenarios of CC and NC cross sections
changing proportionally (left table) and only NC varying and
standard behavior of CC (right table). The results are shown only
for the best cases with both upgoing and downgoing tagged events
and with track vs. shower events discrimination.

%%%%%%%%%%%%%%%%%%%%%%%%%%%%%%%%%%%%%%%%%%%%%%%%%%%%%%%
\section{Discussion and conclusions}\label{conclusions}
%%%%%%%%%%%%%%%%%%%%%%%%%%%%%%%%%%%%%%%%%%%%%%%%%%%%%%%
In this paper we have performed an analysis of the capability of a
km$^3$ NT to disentangle the high energy diffuse neutrino flux
and the  neutrino-nucleon cross section in the PeV energy region
for the incoming neutrinos. As well known, a separation is
possible by exploiting the energy--zenith angular event
distribution. Differently from previous treatments, we have used a
simple phenomenological parametrization for fluxes and cross
sections, presenting a sensitivity analysis in a given parameter
space, rather than for a few benchmark cases. Our forecast is thus
independent from strong assumptions on the normalization or energy
dependence of the flux. The results we obtain are summarized in
Fig.s~\ref {fig2}, \ref{contour1cc}, \ref{contour2cc},
\ref{contour1nc}, \ref{contour2nc}, and refer both to the cases
where CC and NC change proportionally to each other, or the case
where only NC events are affected by new physics.

One important conclusion is that the capability to tag downgoing
neutrino showers in the PeV energy range against the cosmic ray
induced background of penetrating muons appears to be a crucial
requirement to derive meaningful constraints on the cross section.
In this case energy {\it and} zenith angle information on diffuse
neutrino flux event rate in a NT (upgoing vs. downgoing) would
greatly improve the diagnostic power for disentangling cross
section from flux.

Of course, our analysis presents several approximations. In
particular, while we took into account the geometry of the site
(for illustration we considered the \verb"NEMO" site in the
Mediterranean), we did not consider specific information on the
detector geometry, efficiency, etc. Also, we assumed that in muon
and tau tracks the energy loss is deterministic,  rather than
stochastic. A more accurate account of the telescope and of the
energy losses is beyond the scope of this work. Yet, we expect
that it would confirm our qualitative findings, while possibly
worsening the accuracy with  which parameters can be realistically
reconstructed. Since our forecasts are likely more optimistic than
a realistic treatment would show, our conclusion on the importance
of downgoing event tagging is reinforced. In general, the
background due to single energetic muons generated from cosmic
rays and surviving  deep underground drops faster with energy than
the expected neutrino signal, and well above the PeV region the
contamination should be negligible. However, detailed simulation
are required to settle down this issue quantitatively;  an active
veto may help pushing down in energy the region where the
background can be disentangled from the signal on an
event-by-event basis.  For the IceCube telescope, in the PeV
energy range downgoing muons from cosmic ray showers should be
vetoed thanks to the IceTop array. For the Mediterranean km$^3$
neutrino telescopes no final design is available, yet. Given the
importance of the present topic, we believe this is an important
issue to be addressed in the design phase of these telescopes.

Even in the case where no downgoing tagging for neutrinos will be
available for a Mediterranean NT, still its upgoing events at the
PeV would provide important constraints on the flux and cross
section if a combined analysis of its data with IceCube ones is
performed. At very least, it should allow to {\it test} the
hypothesis that the diffuse neutrino fluxes at the PeV from the
two hemispheres are consistent with each other, thus validating
the hypothesis of an overall isotropic flux.

%%%%%%%%%%%%%%%%%
\acknowledgements
%%%%%%%%%%%%%%%%%
P.D. Serpico acknowledges support by the US Department of Energy
and by NASA grant NAG5-10842. G. Miele acknowledges supports by
Generalitat Valenciana (ref. AINV/2007/080 CSIC) and by PRIN 2006
``Fisica Astroparticellare: neutrini ed universo primordiale" by
italian MIUR. S.\ Pastor was supported by the ILIAS project
(contract No.\ RII3-CT-2004-506222), by the Spanish grants
FPA2005-01269 (MEC) and ACOMP07-270 (Generalitat Valenciana), and
by a {\em Ram\'{o}n y Cajal} contract.

%%%%%%%%%%%%%%%%%%%%%%%%%%%

\end{document}